\documentclass[11pt,a4paper,twocolumn]{article}

\usepackage[margin=0.85in]{geometry}
\usepackage[T1]{fontenc}
\usepackage[utf8]{inputenc}
\usepackage{lmodern}
\usepackage{microtype}
\usepackage{placeins} 
\usepackage{tikz}
\usepackage{booktabs}
\usetikzlibrary{positioning, arrows.meta, calc}

\usepackage{authblk}            
\usepackage{graphicx}
\usepackage{booktabs}           
\usepackage{array}
\usepackage{multirow}
\usepackage{xcolor}
\usepackage{amsmath,amssymb,amsfonts}
\usepackage{caption}
\usepackage{subcaption}
\usepackage{float}
\usepackage{enumitem}
\usepackage{url}
\usepackage[hidelinks]{hyperref}
\usepackage{caption}

\newcommand{\benchname}{coRAN-UPF-Bench}
\usepackage{pgfplots}
\pgfplotsset{compat=1.18}

\usepackage{listings}
\usepackage[numbers,sort&compress]{natbib}

\title{\bfseries Comparing UPF Dataplane I/O Modes in a Cloud-Native 5G Core:
\texttt{AF\_PACKET}, \texttt{AF\_XDP}, \texttt{CNDP}, and \texttt{DPDK} on SD-Core BESS-UPF}

\author[1]{Shiva Valia}
\author[1]{Nitin Rajput}
\author[1]{Aditya Gairola}
\author[2]{Vipin Rathi}
\affil[1]{coRAN Labs Private Limited, New Delhi, India}
\affil[2]{Ramanujan College, University of Delhi, New Delhi, India}
\date{}

\begin{document}

\makeatletter
\twocolumn[
  \begin{@twocolumnfalse}
    \maketitle
    \vspace{1em}
  \end{@twocolumnfalse}
]
\makeatother

\begin{abstract}
\noindent The User Plane Function (UPF) carries every byte of
user-plane traffic in a 5G network, handling GTP-U tunneling,
packet detection rules, and QoS enforcement between the gNodeB
and the data network. Its throughput depends heavily on how
packets move between the network interface card and the
user-plane application, which in turn depends on the packet
I/O mode used. Modern open-source UPF implementations support
several such modes, among the most widely used being
\texttt{AF\_PACKET}, \texttt{\texttt{AF\_XDP}}, the Cloud Native Data Plane (CNDP), and
the Data Plane Development Kit (DPDK). Each takes a different
approach to balancing performance, kernel feature reuse, and
ease of deployment in a cloud-native environment.

This paper presents a study comparing these four dataplane
modes on a single open-source UPF. We use SD-Core BESS-UPF,
deployed as a Charmed operator (Juju-orchestrated
\texttt{charmed sd-core} / aether-onramp bundle) on Intel
XXV710 NICs in a Canonical Kubernetes cluster on an x86\_64 platform, and run the same
packet-processing pipeline (GTP-U encap/decap, PDR/FAR/QER
enforcement) under each mode by changing only the BESS port
driver. For each mode
we describe its architecture,
datapath, memory model, and deployment requirements, and we
measure throughput, latency, CPU usage, and stability under
sustained load using \benchname{}, our UPF
benchmarking framework. The deployed UPF is additionally validated
end-to-end against a disaggregated 5G RAN, an OCUDU
7.2-split CU/DU with a Liteon RU and a commercial UE,
confirming it carries real subscriber traffic, not just
synthetic load. We also share the deployment issues we ran
into during real runs and the fixes that resolved them.

On a single Intel XXV710 / \texttt{i40e} testbed
(dual-socket Xeon Gold 6148, RT kernel, 25~GbE SR-IOV with
on-chip VEB hairpin) and the same BESS-UPF pipeline, we
report Non-Drop Rate (NDR) measurements per RFC~2544 / ETSI
NFV-TST~009 across 1--8 datapath workers. Because the
lab optics negotiated 1~GbE, all rates above
$\approx$1.488~Mpps/port are measured VEB-internal
through the on-chip i40e switch and therefore
characterise the UPF's internal processing capacity
rather than wire-forwarded throughput. At 64~B,
CNDP sustains 2.65 / 5.52~Mpps (1w / 2w, 2.09$\times$ scaling),
\texttt{\texttt{AF\_XDP}} sustains 3.69 / 6.47~Mpps (1w / 2w, 1.75$\times$ scaling),
and DPDK (native \texttt{net\_iavf} PMD over \texttt{vfio-pci})
sustains 2.99 / 5.56 / 10.30~Mpps (1w / 2w / 4w, 3.44$\times$
scaling) and 16.0~Mpps bidirectional at 8 workers, with
the lowest latency of the three (8.1~$\mu$s average, 16.2~$\mu$s
P99). \texttt{AF\_PACKET} sits an order of magnitude lower at 0.25~Mpps,
consistent with the kernel-stack cost of every packet
traversal. At matched worker counts the three kernel-bypass modes
are close: at 2 workers, \texttt{AF\_XDP} (6.47~Mpps) leads, with
DPDK (5.56~Mpps) and CNDP (5.52~Mpps) within measurement
noise; per core, \texttt{AF\_XDP} also leads (3.69 versus 2.99 for
DPDK and 2.65 for CNDP). DPDK's absolute lead emerges only
above two workers, where its native PMD over
\texttt{vfio-pci} escapes the per-netdev \texttt{AF\_XDP} socket
limit that pins CNDP and \texttt{AF\_XDP} at two workers, letting
it scale to 10.30~Mpps (4w) and 13.09~Mpps (8w).
A key deployment finding: the original DPDK
result was capped at 6.2~Mpps until we discovered the pod's
Kubernetes \texttt{limits.cpu} was smaller than the datapath
worker count, causing CFS to throttle the polling workers to
$\sim$49\,\% CPU; raising the limit lifted throughput to
10.3~Mpps cleanly.
CNDP and \texttt{\texttt{AF\_XDP}} remain the cloud-native sweet spot when
\texttt{vfio-pci} binding, hugepages, and isolated cores are
unaffordable; DPDK is justified where the operator owns the
NIC and can absorb the deployment footprint. The
paper provides a side-by-side reference for operators choosing a
UPF dataplane mode in a cloud-native 5G deployment.

\end{abstract}

\vspace{0.6em}
\texttt{AF\_PACKET}, \texttt{\texttt{AF\_XDP}}, CNDP, DPDK, GTP-U, Packet I/O, Cloud-Native
5G Core, Kernel Bypass, Packet Processing.

\section{Introduction}
\label{sec:intro}

The User Plane Function (UPF) is a core component of every 5G
network. As specified in 3GPP TS~23.501~\cite{ts23501}, it
anchors PDU sessions, enforces per-packet rules in the form of
Packet Detection Rules (PDRs), Forwarding Action Rules (FARs),
QoS Enforcement Rules (QERs), and Usage Reporting Rules (URRs),
and tunnels user traffic between the gNodeB and the data
network using GTP-U over UDP~\cite{ts29281}. Every byte that a subscriber
sends or receives passes through the UPF. This makes its
throughput a direct factor in how many users a single server
can handle, and therefore in the cost of running the network.

How the UPF actually moves packets between the network
interface card and the application is the main lever for that
throughput. The Linux kernel and the open-source ecosystem
provide several ways to do this. Standard kernel networking
uses \texttt{AF\_PACKET}, the same socket family that tools like
\texttt{tcpdump} rely on. \texttt{\texttt{AF\_XDP}} uses eBPF and XDP sockets to
bypass most of the kernel stack while still keeping the kernel
driver. The Cloud Native Data Plane (CNDP) builds a userspace
abstraction on top of \texttt{\texttt{AF\_XDP}} and offers a DPDK-like
programming model without DPDK's heavier requirements. The
Data Plane Development Kit (DPDK) goes the furthest, taking
the NIC fully into userspace through a Poll-Mode Driver.

Each of these modes is well-described on its own, and several
papers compare them in pairs (most often \texttt{\texttt{AF\_XDP}} against DPDK) on synthetic packet generators. What is missing is a
side-by-side look at all four on the same UPF pipeline, under
the same workload, with a real 5G control plane in the loop.
Operators choosing a UPF dataplane today often have to combine
results from different testbeds and different applications to
make a decision.

This paper presents that comparison. We deploy the open-source
SD-Core BESS-UPF on an Intel XXV710 NIC as a
\emph{Charmed operator}~\cite{juju-charms} (the upstream
\texttt{charmed sd-core} bundle, orchestrated by Juju on top of Canonical Kubernetes),
and we run the same BESS packet-processing pipeline (GTP-U encap/decap and PDR/FAR/QER enforcement) under all four I/O modes.
The only thing that changes between runs is the BESS
port driver. For each mode we walk through its architecture,
datapath, memory model, and deployment requirements, and we
report throughput, latency and jitter, CPU and core scaling,
and stability under sustained load. We also document the
operational problems we ran into during real deployments and
the fixes that brought each mode to a stable state.

\subsection{Contributions}
\label{sec:contributions}

This paper makes the following contributions:

\begin{itemize}[leftmargin=*]
  \item A side-by-side comparison of four widely-used UPF
        packet I/O modes (\texttt{AF\_PACKET}, \texttt{\texttt{AF\_XDP}}, CNDP, and DPDK) on a single
        UPF pipeline, with the I/O backend
        as the only variable.
  \item Architectural walkthroughs of each mode, covering
        datapath flow, memory model, execution model, and
        deployment requirements.
  \item Benchmark results from a real testbed
  (Intel XXV710/\texttt{i40e} on x86\_64, Canonical Kubernetes,
   charmed sd-core / aether-onramp) covering
        throughput, latency and jitter, CPU and core scaling,
        and stability under sustained load.
  \item Operational findings from running each mode in
        production, including the failure modes we encountered
        and the fixes that resolved them.
  \item A short decision guide for operators on which mode
        fits which deployment.
  \item Deployment of the UPF as a Charmed operator
        (Juju-orchestrated) within the charmed sd-core
        bundle on Kubernetes, demonstrating that the
        cloud-native dataplane comparison holds under a
        production-grade operator lifecycle (rolling
        upgrades, declarative config, atomic relations).
  \item End-to-end functional validation against a
        disaggregated 5G RAN with an OCUDU 7.2-split CU/DU pair,
        a Liteon Open RAN radio unit (RU), and a commercial UE, exercising
        the deployed UPF
        through the full UE $\rightarrow$ gNB
        $\rightarrow$ UPF $\rightarrow$ data-network path.
  \item \benchname{}, our UPF benchmarking
        framework built to drive every measurement in this
        paper. It orchestrates synthetic load generation,
        in-pipeline BESS probe insertion, soak runs, and
        drop/forwarding-efficiency reconciliation into a
        single mode-agnostic harness over the four
        dataplane backends.
\end{itemize}

\subsection{Paper Organization}
\label{sec:org}

Section~\ref{sec:background} introduces the 5G UPF, the Linux
packet I/O primitives, and the BESS port abstraction that makes
the four modes interchangeable. Section~\ref{sec:modes} walks
through each of the four dataplane modes in turn.
Section~\ref{sec:compare} brings the architectural differences
together in a single comparison.
Section~\ref{sec:method} describes the testbed and methodology,
and Section~\ref{sec:results} presents the benchmark results.
Section~\ref{sec:ops} reports the deployment pitfalls we hit
and the fixes that resolved them.
Section~\ref{sec:related} situates the work against prior
art, and Section~\ref{sec:discuss} translates the
comparison into a decision framework for operators.
Section~\ref{sec:future} discusses limitations and future
work, and Section~\ref{sec:conclusion} concludes.

\section{Background}
\label{sec:background}

This section sets up the three pieces of vocabulary the rest of
the paper depends on: the 5G UPF and its user-plane interfaces
(\S\ref{sec:bg-upf}), the Linux packet I/O primitives that all
four modes are built on (\S\ref{sec:bg-io}), and the BESS port
abstraction that makes the four modes interchangeable on a
single pipeline (\S\ref{sec:bg-bess}).

\subsection{The 5G UPF and the User Plane}
\label{sec:bg-upf}

The UPF is part of the 5G Core's Service Based Architecture and
handles all user-plane traffic for a 5G network. The control
plane sits separately, with the Session Management Function
(SMF) instructing the UPF over the N4 interface using the
Packet Forwarding Control Protocol (PFCP), defined in 3GPP
TS~29.244~\cite{ts29244}. The UPF's main interfaces are:

\begin{itemize}[leftmargin=*]
  \item \textbf{N3:} connects the gNodeB to the UPF, carrying
        user data inside a GTP-U tunnel over UDP. Each tunnel
        is identified by a Tunnel Endpoint Identifier (TEID),
        and the inner packet is the actual IP packet for the
        user equipment.
  \item \textbf{N6:} connects the UPF to the data network (the
        Internet or a private network), carrying the
        decapsulated user traffic.
  \item \textbf{N9:} connects two UPFs when traffic chains
        through more than one UPF.
\end{itemize}

For every PDU session, the SMF programs the UPF with four
classes of rules:

\begin{itemize}[leftmargin=*]
  \item \textbf{Packet Detection Rules (PDRs)} match an
        incoming packet against fields such as the TEID, source
        and destination IP and port, and QoS Flow Identifier.
  \item \textbf{Forwarding Action Rules (FARs)} describe what
        do with a matched packet: forward, drop, buffer, or encapsulate.
  \item \textbf{QoS Enforcement Rules (QERs)} apply per-flow
        rate limiting.
  \item \textbf{Usage Reporting Rules (URRs)} drive charging
        and quota.
\end{itemize}

Every user packet, in both directions, is matched against the
PDR table, mutated by its FAR (typically a GTP-U encap or decap
and a routing decision), and metered against its QER, all at
line rate. This per-packet rule processing is the UPF's main
computational workload, and the I/O backend is what feeds it.

Several open-source UPF implementations exist. The Aether
project's SD-Core distribution~\cite{sdcore-upf}, which we use
in this paper, ships a UPF built on BESS, the Berkeley
Extensible Software Switch~\cite{han2015bess}, and supports
multiple packet I/O backends through BESS's port abstraction
(\S\ref{sec:bg-bess}). Other open-source UPFs include
free5GC~\cite{free5gc}, which has both a kernel-module
(\texttt{gtp5g}) datapath and a DPDK-based userspace datapath,
and eUPF~\cite{eupf}, an eBPF-only UPF that runs entirely at
the XDP hook. Each project picks a different default I/O mode;
this paper holds the UPF (SD-Core BESS-UPF) constant and varies
the I/O.

\subsection{Linux Packet I/O Primitives}
\label{sec:bg-io}

Every dataplane mode in this paper is, at some level, a
different answer to the same question: where in the Linux
packet I/O pipeline do we intercept the packet, and what do we
hand to the application? A short tour of the underlying
primitives makes the later per-mode sections easier to follow.

\paragraph{NIC and descriptor rings.}
A modern NIC exposes one or more queue pairs, each made up of
an RX (receive) and TX (transmit) descriptor ring in host
memory. Each descriptor is a small fixed-size record (typically
16--32 bytes) pointing at a packet buffer and carrying packet
metadata. The NIC writes incoming packets into the buffers via
DMA, advances a hardware head pointer, and signals the host
with an MSI-X interrupt.

\begin{figure}[H]
    \centering
    \includegraphics[width=\columnwidth]{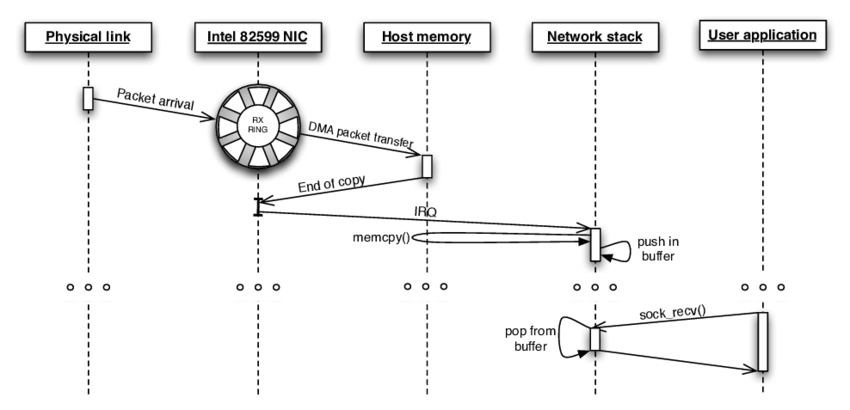}
    \caption{RX descriptor ring structure showing the circular buffer, descriptors, head and tail pointers, and DMA packet flow between the NIC and host memory (adapted from \cite{tungdam_ringbuffers}).}
    \label{fig:nic-descriptor-ring}
\end{figure}
\paragraph{Interrupts, NAPI, and the kernel stack.}
In the standard Linux datapath, an MSI-X interrupt wakes the  NIC driver, which enters NAPI~\cite{napi} (a hybrid
interrupt-plus-poll scheme designed to avoid livelock under
heavy load). The driver
pulls descriptors from the RX ring in batches and processes
incoming packets using NAPI polling rather than handling every
packet through a separate interrupt.
The driver then allocates a kernel socket buffer
(\texttt{sk\_buff}) for each packet and hands the chain to the
network stack. The stack performs L2/L3/L4 demultiplexing, runs
netfilter and traffic-control hooks, and finally puts the
packet on a destination socket. A userspace application reading
from that socket pays one extra copy from kernel space to user
space per packet. The combined cost of \texttt{sk\_buff}
allocation, stack traversal, and the copy-to-user is what
kernel-bypass techniques try to eliminate.
\begin{figure}[H]
    \centering
    \includegraphics[width=\columnwidth]{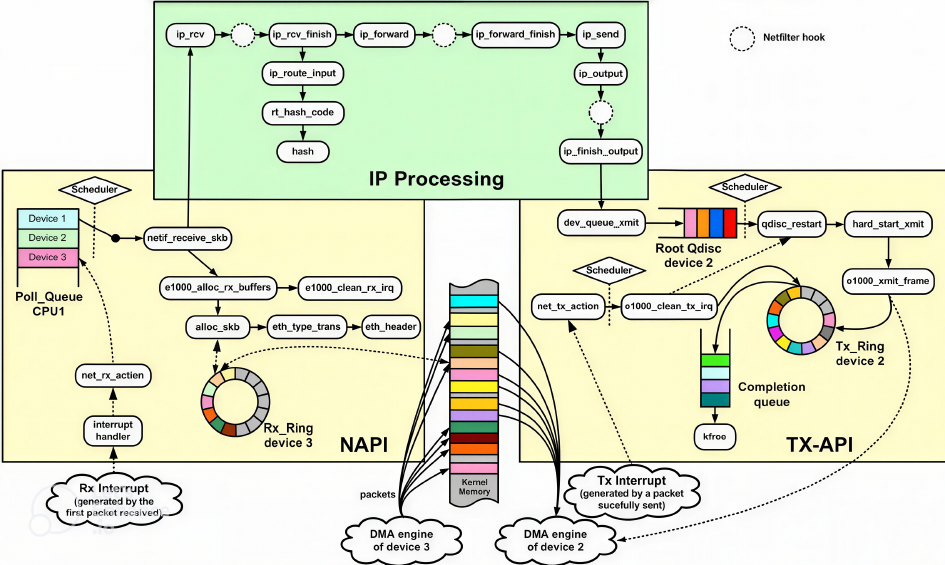}
    \caption{Linux NAPI packet processing flow illustrating interrupt mitigation through hybrid interrupt-driven activation and polling-based packet processing (adapted from \cite{boutnaru_napi}).}
    \label{fig:napi-flow}
\end{figure}

\paragraph{Zero-copy and shared memory.}
Zero-copy packet I/O removes the kernel-to-user memcpy by
exposing the NIC's DMA buffers directly to the application,
usually through a shared-memory region. The buffer is filled by
the NIC, processed in place by the application, and returned to the NIC, 
with no copy and no \texttt{sk\_buff}. \texttt{\texttt{AF\_XDP}} achieves
this through its UMEM construct
(\S\ref{sec:mode-afxdp}); DPDK achieves it through its
\texttt{mbuf} pool over hugepages
(\S\ref{sec:mode-dpdk}).

\paragraph{Hugepages.}
Conventional Linux pages are 4~KB. A typical packet buffer pool
of tens of thousands of packets quickly exceeds the Translation
Lookaside Buffer (TLB) capacity, which causes TLB misses on
every buffer access. Hugepages (2~MB or 1~GB on x86) reduce TLB
pressure by several orders of magnitude. DPDK requires
hugepages; \texttt{\texttt{AF\_XDP}} can use them as an optimization but does
not require them.

\paragraph{VFIO and IOMMU.}
Userspace direct DMA needs both safety (the application must
not be able to DMA into arbitrary kernel memory) and isolation
(one tenant must not be able to DMA into another's). The IOMMU
provides the hardware boundary, and Linux's VFIO subsystem
(\texttt{vfio-pci}) exposes a NIC to a privileged userspace
process as a member of an IOMMU group. DPDK depends on VFIO;
\texttt{\texttt{AF\_XDP}} and CNDP do not.

\paragraph{Poll-mode vs.\ interrupt-mode.}
An interrupt-driven datapath is event-triggered: the CPU is
idle until the NIC raises an interrupt. A poll-mode datapath
dedicates one or more CPU cores to spinning on the RX ring,
trading CPU cycles for sub-microsecond reaction times.
\texttt{AF\_PACKET} is interrupt-driven, DPDK is poll-mode, and
\texttt{\texttt{AF\_XDP}} and CNDP can be configured either way through their
\texttt{busy\_poll} setting.

\paragraph{RSS and multi-queue.}
A single CPU core can only process so many packets per second.
All four modes scale horizontally by binding multiple worker
threads to multiple NIC RX queues, with incoming packets
steered across queues by Receive-Side Scaling (RSS), 
which is a hash over selected packet fields. This works well for plain IP/TCP
traffic but is less straightforward for GTP-U, since the
default RSS hash does not look at the inner packet's headers.

\subsection{BESS-UPF and the Port Abstraction}
\label{sec:bg-bess}

The Berkeley Extensible Software Switch
(BESS)~\cite{han2015bess} is a programmable soft-switch that
represents a packet-processing pipeline as a directed graph of
modules. Each module performs a small, well-defined operation: match
against a table, encap or decap a header, drop, meter, or
count. Packets traverse the graph in batches for
cache efficiency. SD-Core's BESS-UPF builds the 5G UPF
datapath as such a graph: a \texttt{PortInc} module reads
batches from the N3 interface, a \texttt{GtpuDecap} module
strips the GTP-U header and recovers the inner IP packet, a
\texttt{WildcardMatch} module implements the PDR table, FAR
actions are applied by a chain of mutators, a metering module
enforces QER rate limits, and a \texttt{PortOut} module writes
the packet to the N6 interface. The symmetric path handles the
return direction.

The part that matters for this paper is BESS's
\textbf{port abstraction}. A BESS port is an endpoint that
exposes \texttt{RecvPackets} and \texttt{SendPackets}
primitives, and everything below those two methods (driver,
ring, memory model, kernel involvement) is encapsulated by
the port type. BESS-UPF picks one of several port types based
on configuration:

\begin{itemize}[leftmargin=*]
  \item \textbf{PMDPort with \texttt{net\_\texttt{af\_packet}}} for the
        \texttt{AF\_PACKET} mode, DPDK's poll-mode driver wrapped
        around a Linux \texttt{AF\_PACKET} raw socket.
  \item \textbf{PMDPort with \texttt{net\_\texttt{\texttt{af\_xdp}}}} for the
        \texttt{\texttt{AF\_XDP}} mode, DPDK's PMD wrapped around an XDP
        socket.
  \item \textbf{CNDPPort} for the CNDP mode, a separate port
        type that calls Intel's CNDP library directly.
  \item \textbf{PMDPort with a native DPDK driver}
        (\texttt{net\_iavf}, \texttt{net\_i40e},
        \texttt{net\_ice}) for the DPDK mode, the NIC is
        bound to \texttt{vfio-pci} and accessed by the
        corresponding PMD.
\end{itemize}

This is what makes the experiment in this paper possible. The
BESS module graph, the PFCP/PDR tables, the GTP-U logic, the
metering rules, everything above \texttt{PortInc} and
\texttt{PortOut}, stays bit-identical across all four modes.
The only thing that changes is which port driver fills and
drains those two endpoints. Any performance difference we
observe is therefore attributable to the I/O backend rather
than to a difference in the pipeline.

\begin{figure*}[t]
  \centering
  \includegraphics[width=\textwidth]{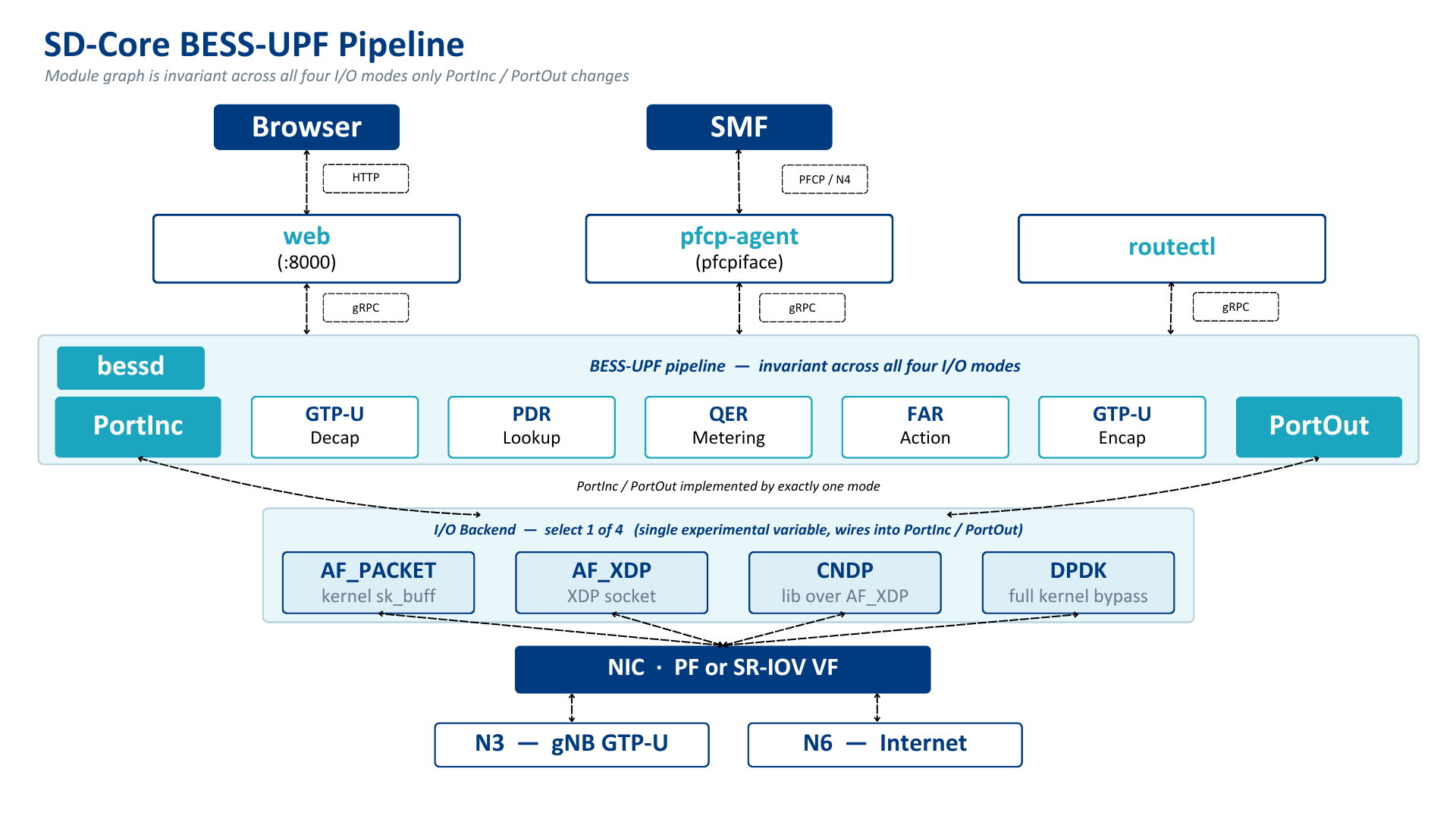}
  \caption{The SD-Core BESS-UPF pipeline. The module graph
    above the dashed line (PDR/FAR/QER processing, GTP-U
    encap/decap) is invariant across all four dataplane modes.
    Only the \texttt{PortInc}/\texttt{PortOut} port driver
    changes, isolating the I/O backend as the single
    experimental variable.}
  \label{fig:bess-pipeline}
\end{figure*}

\section{The Four Dataplane Modes}
\label{sec:modes}

This section walks through the four dataplane modes one at a
time. Each mode is described using the same seven-point
structure (overview, architecture, datapath packet flow,
memory model, deployment requirements, advantages and
limitations, and key findings) so that the modes can be
compared like-for-like in Section~\ref{sec:compare}. We present
them in order of increasing kernel bypass, from \texttt{AF\_PACKET}
(full kernel stack) to DPDK (full kernel bypass), with \texttt{\texttt{AF\_XDP}}
and CNDP in between.

\subsection{\texttt{AF\_PACKET} (Static Kernel Module Mode)}
\label{sec:mode-afpacket}

\paragraph{Overview.}
\texttt{AF\_PACKET} is the standard Linux raw-socket family used by
tools like \texttt{tcpdump} and Wireshark. It is the oldest and
most compatible of the four modes: the packet travels through
the full Linux network stack, and the application reads it
from a raw socket. In BESS-UPF, \texttt{AF\_PACKET} is exposed through
DPDK's \texttt{net\_\texttt{af\_packet}} poll-mode driver, which wraps
a Linux \texttt{PF\_PACKET} socket behind DPDK's port API. No
special NIC driver, no hugepages, no VFIO, and no kernel
patching are required, which makes this mode the easiest to
deploy in a generic container environment. We use it as the
baseline against which the three faster modes are compared.

\begin{figure}[t]
\centering
\resizebox{\columnwidth}{!}{%
\begin{tikzpicture}[
  font=\small,
  >=Latex,
  every node/.style={align=center},
  layer/.style={
    rectangle, draw=black!70, rounded corners=2pt, thick,
    minimum width=4.8mm, minimum height=9mm, inner sep=4pt
  },
  app/.style   ={layer, fill=orange!25, draw=orange!70!black},
  user/.style  ={layer, fill=cyan!18,   draw=cyan!60!black},
  kernel/.style={layer, fill=violet!18, draw=violet!60!black},
  hw/.style    ={layer, fill=gray!30,   draw=gray!70},
  cost/.style  ={font=\scriptsize\itshape, text=gray!50!black, align=left}
]

\node[app]                        (bess)
  {BESS-UPF Pipeline\\
   \scriptsize PortInc $\rightarrow$ GTP-U decap $\rightarrow$ PDR/FAR/QER $\rightarrow$ PortOut};
\node[user,   below=3mm of bess]  (pmd)
  {\texttt{net\_\texttt{af\_packet}} PMD\\
   \scriptsize DPDK port wrapping a raw socket};
\node[kernel, below=8mm of pmd]   (sock)
  {\texttt{PF\_PACKET} Raw Socket\\
   \scriptsize \texttt{recvmsg()} / optional \texttt{TPACKET\_V3} ring};
\node[kernel, below=3mm of sock]  (stack)
  {Linux Network Stack\\
   \scriptsize L2/L3/L4 demux, netfilter, tc};
\node[kernel, below=3mm of stack] (napi)
  {NIC Driver + NAPI\\
   \scriptsize \texttt{sk\_buff} allocation per packet};
\node[hw,     below=8mm of napi]  (nic)
  {Network Interface Card\\
   \scriptsize RX / TX descriptor rings};

\draw[<->, thick] (bess)  -- (pmd);
\draw[<->, thick] (pmd)   -- (sock);
\draw[<->, thick] (sock)  -- (stack);
\draw[<->, thick] (stack) -- (napi);
\draw[<->, thick] (napi)  -- (nic);

\path (pmd.south) -- (sock.north) coordinate[midway] (b1);
\draw[dashed, thick, gray]
  ($(b1)+(-3.2,0)$) -- ($(b1)+(3.2,0)$)
  node[right=1mm, font=\scriptsize, text=gray!60!black] {user $|$ kernel};

\path (napi.south) -- (nic.north) coordinate[midway] (b2);
\draw[dashed, thick, gray]
  ($(b2)+(-3.2,0)$) -- ($(b2)+(3.2,0)$)
  node[right=1mm, font=\scriptsize, text=gray!60!black] {kernel $|$ NIC};

\node[cost, anchor=east] at ($(pmd.west)   +(-5mm,0)$) {copy to/from\\user space};
\node[cost, anchor=east] at ($(stack.west) +(-5mm,0)$) {full stack\\traversal};
\node[cost, anchor=east] at ($(napi.west)  +(-5mm,0)$) {\texttt{sk\_buff}\\per packet};
\node[cost, anchor=east] at ($(nic.west)   +(-5mm,0)$) {DMA +\\MSI-X IRQ};

\end{tikzpicture}%
}
\caption{\texttt{AF\_PACKET} datapath in SD-Core BESS-UPF. Packets
traverse the full Linux network stack: NAPI allocates an
\texttt{sk\_buff} for each packet, the stack performs L2/L3/L4
demultiplexing (plus any active netfilter and \texttt{tc}
hooks), the kernel clones the packet onto a
\texttt{PF\_PACKET} raw socket, and the
\texttt{net\_\texttt{af\_packet}} DPDK PMD copies it to user space for
the BESS pipeline. The transmit path is symmetric. No
hugepages, VFIO, or kernel-bypass drivers are required.}
\label{fig:afpacket-arch}
\end{figure}
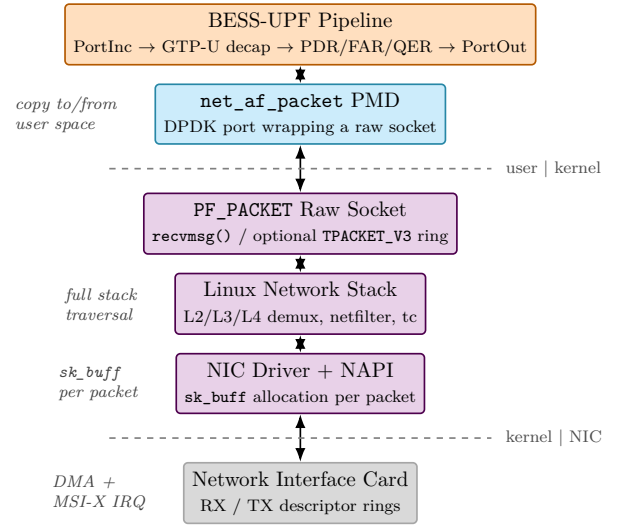

\paragraph{Architecture.}
Figure~\ref{fig:afpacket-arch} shows the \texttt{AF\_PACKET} datapath.
A packet arriving at the NIC is processed by the kernel NIC
driver, passed up through the standard network stack
(NAPI~$\rightarrow$ \texttt{sk\_buff}~$\rightarrow$ IP/UDP
demultiplexing), and finally placed on a \texttt{PF\_PACKET}
raw socket bound to the interface. BESS's PortInc module, via
the \texttt{net\_\texttt{af\_packet}} PMD, reads from this socket using
\texttt{recvmsg()} (or the \texttt{TPACKET\_V3} ring-buffer
interface if available). The packet is then handed to the BESS
module graph for GTP-U decap, PDR/FAR/QER processing, and
forwarding. The reverse path on transmit goes from the BESS
\texttt{PortOut} module back into the kernel through the same
raw socket, then out through the standard stack to the NIC.

\paragraph{Datapath packet flow.}
On receive, a single packet traverses the following stages:

\begin{enumerate}[leftmargin=*]
  \item The NIC writes the packet into an RX descriptor ring
        via DMA and raises an MSI-X interrupt.
  \item The kernel NIC driver enters NAPI, pulls a batch of
        descriptors, and allocates an \texttt{sk\_buff} for
        each.
  \item The packet is pushed up through the network stack: L2
        demultiplexing, IP/UDP protocol handlers, netfilter
        and traffic-control hooks.
  \item Because the \texttt{AF\_PACKET} socket is bound to the
        interface in promiscuous (or directed) mode, the kernel
        clones the packet to the socket's receive queue.
  \item The application (BESS's PMD, in our case) reads the
        packet from the socket. This involves a copy from
        kernel space to user space.
  \item The packet enters BESS's module graph through the
        \texttt{PortInc} module and is processed by the rest of
        the UPF pipeline.
\end{enumerate}

The transmit path is symmetric: BESS writes the packet to the
socket, the kernel allocates an \texttt{sk\_buff}, and the
packet is pushed back down through the stack and out the NIC.

\paragraph{Memory model.}
\texttt{AF\_PACKET} uses the kernel's \texttt{sk\_buff} for all packet
representation. There is one \texttt{sk\_buff} allocation per
packet (typically from a slab cache), and at least one memory
copy on every receive (kernel space~$\rightarrow$ user space).
The \texttt{TPACKET\_V3} interface can reduce per-packet
overhead by exposing a ring buffer of fixed-size frame slots,
but the underlying \texttt{sk\_buff} allocation and kernel
stack traversal are unchanged. No hugepages, no dedicated
memory pools, and no IOMMU mappings are involved.

\paragraph{Deployment requirements.}
\texttt{AF\_PACKET} is the most portable of the four modes. The
requirements are:

\begin{itemize}[leftmargin=*]
  \item \textbf{NIC driver:} any Linux-supported NIC. No XDP
        support, no DPDK PMD, and no VFIO binding are needed.
        The interface stays under the kernel.
  \item \textbf{Kernel features:} a stock Linux kernel.
        \texttt{CAP\_NET\_RAW} is required to open a
        \texttt{PF\_PACKET} socket; in a Kubernetes pod this
        is granted via the security context.
  \item \textbf{Hugepages:} not required.
  \item \textbf{Memlock and privileged mode:} not required.
        The pod can run with the default security profile in
        most clusters.
  \item \textbf{CNI:} any CNI works, since the NIC is owned by
        the kernel. We use the standard Calico/host-device
        setup with no extra plumbing.
\end{itemize}

\paragraph{Advantages and Limitations.}
\textit{Advantages.}
\texttt{AF\_PACKET} is the easiest mode to deploy in a generic
cloud-native environment: no kernel tuning, no hardware
constraints, no privileged containers. Standard kernel features
(\texttt{tc}, \texttt{iptables}, netfilter, eBPF tracing, etc.)
remain available because the packet still traverses the full
stack. This makes the mode useful for development and
debugging, and for deployments where compatibility matters
more than raw throughput.

\textit{Limitations.}
The same property that makes \texttt{AF\_PACKET} compatible is what
makes it slow. Every packet pays for an \texttt{sk\_buff}
allocation, full stack traversal, and a kernel-to-user copy.
CPU usage per packet is the highest of the four modes, and
the per-core packets-per-second ceiling is correspondingly the
lowest. Multi-queue/RSS scaling helps but cannot close the gap
with the kernel-bypass modes.

\paragraph{Key findings.}
\texttt{AF\_PACKET} serves as a stable baseline. Throughput is bounded
primarily by per-packet CPU cost in the kernel stack, not by
any BESS-side bottleneck. We observe that:

\begin{itemize}[leftmargin=*]
        \item \textbf{Throughput.} \texttt{AF\_PACKET} sustains
        0.252~Mpps full-pipeline at 86~B with two workers
        (0.224~Mpps single-worker), the lowest of the four
        modes by roughly an order of magnitude, bounded by
        per-packet \texttt{sk\_buff} allocation and the
        kernel-to-user copy. The rate is essentially flat
        across frame sizes (0.252 $\to$ 0.233~Mpps from 86~B
        to 1280~B), confirming the pps-bound character of the
        kernel-stack path.

  \item CPU utilization climbs steeply with offered load, with
        the kernel softirq path dominating perf profiles.
  \item Deployment is the simplest of the four modes: no
        hugepages, no VFIO, no privileged container, no
        special CNI.
  \item Latency under light load is comparable to \texttt{\texttt{AF\_XDP}},
        because the kernel stack is fast in the uncontended
        case; the gap appears under sustained load.
\end{itemize}

This makes \texttt{AF\_PACKET} a reasonable choice for low-throughput
edge deployments, lab testbeds, and any environment where the
operational simplicity outweighs the throughput penalty.

\FloatBarrier

\subsection{\texttt{\texttt{AF\_XDP}}}
\label{sec:mode-afxdp}

\paragraph{Overview.}
\texttt{\texttt{AF\_XDP}} is a Linux socket family (\texttt{PF\_XDP}) introduced
in kernel 4.18 and matured through the 5.x series, designed
for high-performance packet I/O without leaving the kernel.
Packets arriving at the NIC are intercepted by an
eBPF~\cite{ebpf} program attached at the eXpress Data Path
(XDP)~\cite{xdp2018} hook, a point in the NIC driver
\emph{before} the kernel allocates an \texttt{sk\_buff} or
runs the rest of the network stack. The XDP program redirects
the packet to an XDP socket (XSK), which exposes it directly
to userspace through a shared, memory-mapped region called
UMEM. On drivers that support it, the entire datapath runs in
zero-copy mode: the NIC writes packets via DMA into UMEM
frames that the application reads in place, with no
\texttt{sk\_buff} allocation and no kernel-to-user copy.

In BESS-UPF, \texttt{\texttt{AF\_XDP}} is exposed through DPDK's
\texttt{net\_\texttt{\texttt{af\_xdp}}} poll-mode driver, which wraps an XDP
socket behind the same port API used by \texttt{AF\_PACKET}. Switching
from \texttt{AF\_PACKET} to \texttt{\texttt{AF\_XDP}} is therefore a configuration change
in the BESS port type, the BESS module graph above
\texttt{PortInc} and \texttt{PortOut} stays bit-identical.

\begin{figure*}[t]
  \centering
  \includegraphics[width=\textwidth]{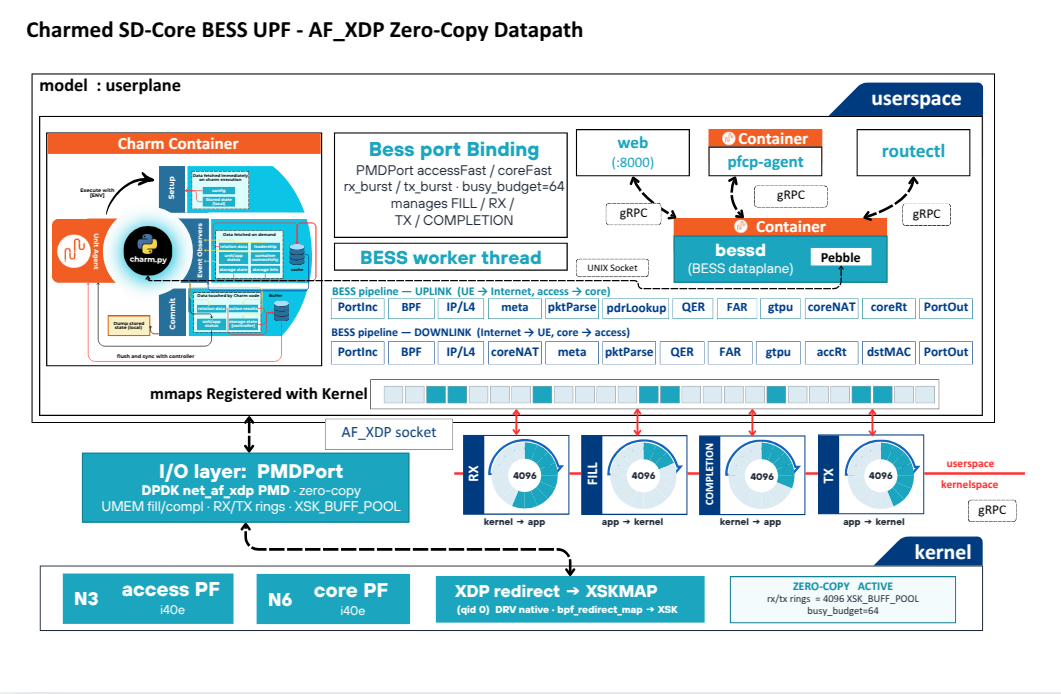}
  \caption{\texttt{\texttt{AF\_XDP}} datapath in the Charmed SD-Core BESS-UPF
    deployment. A single BESS worker (DPDK
    \texttt{net\_\texttt{\texttt{af\_xdp}}} PMD with \texttt{busy\_budget=64})
binds an \texttt{\texttt{AF\_XDP}} socket against the \texttt{i40e} access
    and core PFs (N3 / N6). Four ring queues
    (RX / Fill / Completion / TX, depth 4096) are
    \texttt{mmap}'d into the BESS worker's address space; the
    kernel-side XDP program redirects incoming packets to
    the XSK via the XSKMAP, and zero-copy is confirmed
    active (\texttt{XSK\_BUFF\_POOL}). The Charm Container
    on the left manages lifecycle.}
  \label{fig:afxdp-arch}
\end{figure*}

\paragraph{Architecture.}
Figure~\ref{fig:afxdp-arch} shows the \texttt{\texttt{AF\_XDP}} datapath as
deployed in our Charmed BESS-UPF (single BESS worker on the
\texttt{net\_\texttt{\texttt{af\_xdp}}} PMD against \texttt{i40e} PFs with
\texttt{busy\_budget=64}). An XDP program (eBPF bytecode
loaded into the NIC driver) runs
on every received packet before any \texttt{sk\_buff} is
allocated. The program returns one of several actions:
\texttt{XDP\_PASS} (forward up the stack), \texttt{XDP\_DROP}
(discard), \texttt{XDP\_TX} (bounce back out), or
\texttt{XDP\_REDIRECT}, which steers the packet to an XSK
listed in an eBPF map. The XSK is bound to a specific
(\textit{netdev}, queue index) pair, so multi-queue NICs scale
horizontally by binding one XSK per RX queue.

Each XSK exposes four ring queues that the application and
the kernel share through \texttt{mmap()}: the RX ring carries
received packet descriptors to the application, the TX ring
carries outgoing packet descriptors to the kernel, the Fill
Queue (FQ) returns empty UMEM frames to the kernel for use as
incoming receive buffers, and the Completion Queue (CQ)
returns transmitted frames to the application. Frames live
in UMEM, an mmap'd contiguous region divided into fixed-size
chunks (typically 2~KB).

\paragraph{Datapath packet flow.}
On receive, a single packet traverses the following stages:

\begin{enumerate}[leftmargin=*]
  \item The NIC writes the packet into a buffer whose UMEM
        address the driver has already taken from the Fill
        Queue. DMA goes directly into UMEM.
  \item The NIC raises an MSI-X interrupt (or the
        application is in busy-poll mode); the driver's XDP
        hook runs the attached eBPF program.
  \item The program returns \texttt{XDP\_REDIRECT}, with the
        target XSK looked up from an eBPF map keyed by the
        RX queue index.
  \item The kernel writes the frame's UMEM offset and length
        into the XSK's RX ring.
  \item The application (the BESS \texttt{net\_\texttt{\texttt{af\_xdp}}} PMD)
        polls the RX ring, reads the descriptor, and
        accesses the packet directly in UMEM. No copy.
  \item The packet enters BESS's module graph through
        \texttt{PortInc} and is processed by the rest of the
        UPF pipeline.
  \item When the application is done with the frame, it
        returns the UMEM address to the Fill Queue so the
        driver can reuse it for another incoming packet.
\end{enumerate}

On transmit, the application places a descriptor (UMEM
offset + length) on the TX ring, the kernel hands it to the
driver for DMA, and the driver writes the completion back to
the Completion Queue once the frame has been sent. Both
directions stay inside the kernel's NIC driver, only the
rest of the network stack is bypassed.

\begin{figure*}[t]
  \centering
  \begin{subfigure}{0.48\textwidth}
    \centering
    \includegraphics[width=\linewidth]{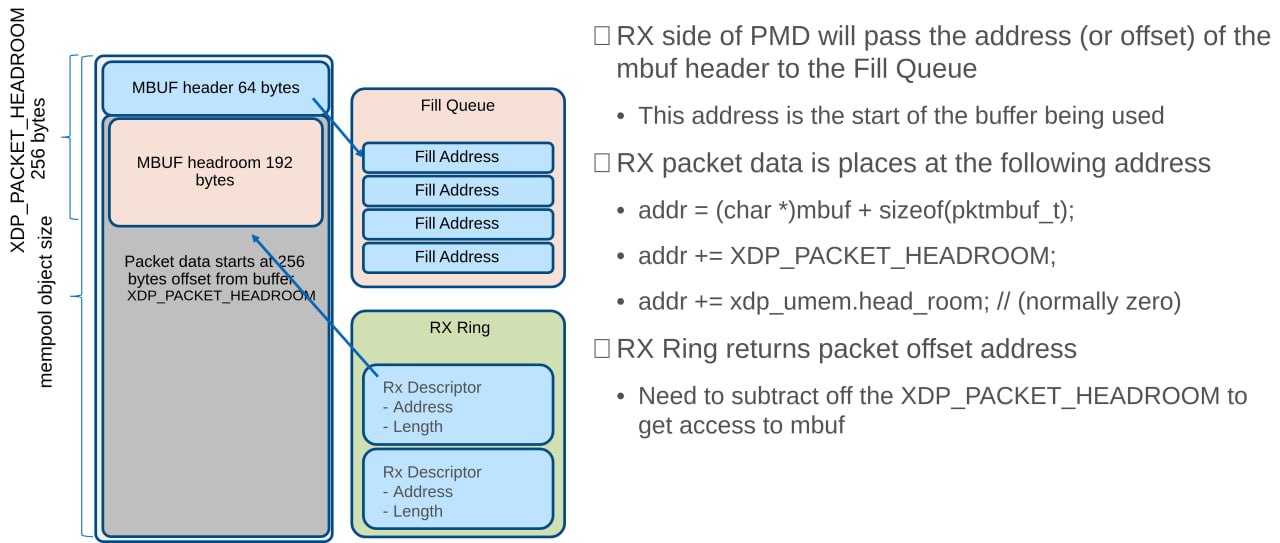}
    \caption{RX side: Fill Queue $\rightarrow$ RX Ring.}
    \label{fig:afxdp-mem-rx}
  \end{subfigure}
  \hfill
  \begin{subfigure}{0.48\textwidth}
    \centering
    \includegraphics[width=\linewidth]{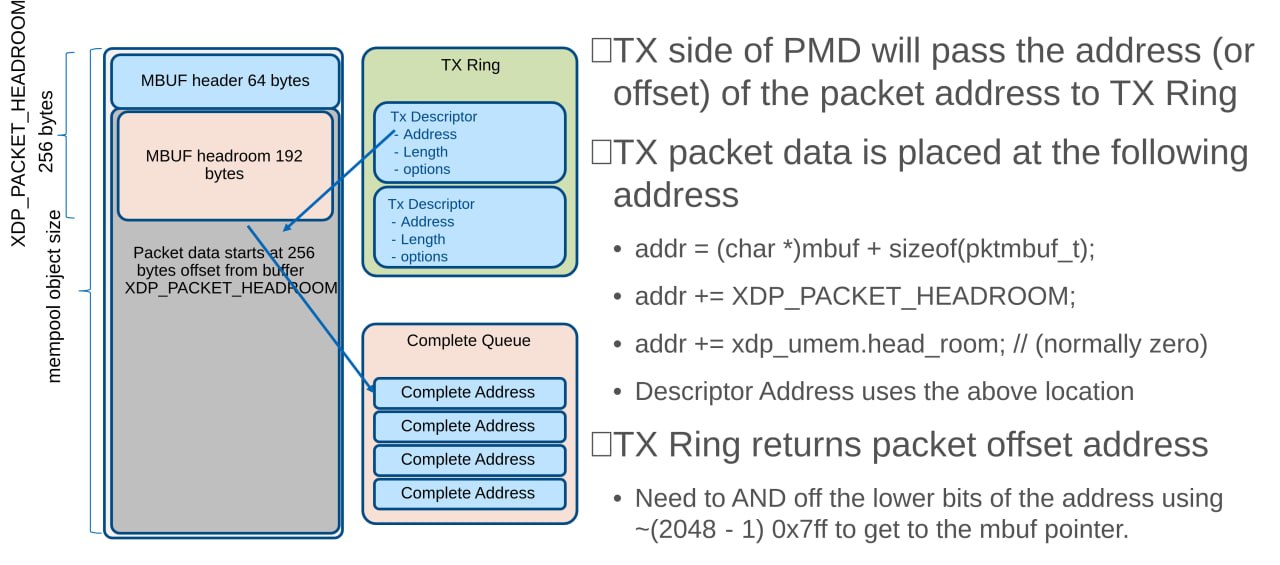}
    \caption{TX side: TX Ring $\rightarrow$ Completion Queue.}
    \label{fig:afxdp-mem-tx}
  \end{subfigure}
  \caption{\texttt{\texttt{AF\_XDP}} UMEM memory layout. The first 256 bytes
    of each UMEM frame (\texttt{XDP\_PACKET\_HEADROOM})
    contain the MBUF header and headroom; the packet payload
    begins after the offset. RX and TX use mirror-image ring
    structures over the same UMEM region. \emph{Source: adapted from~\cite{intel-afxdp-guide}.}}
  \label{fig:afxdp-mem}
\end{figure*}

\paragraph{Memory model.}
\texttt{\texttt{AF\_XDP}}'s central abstraction is the \textbf{UMEM}: a
contiguous, memory-mapped region of userspace memory that
both the application and the kernel can access. UMEM is
divided into fixed-size frames (typically 2~KB), and every
packet, received or transmitted, lives in a UMEM frame.
Figures~\ref{fig:afxdp-mem-rx} and~\ref{fig:afxdp-mem-tx}
show the layout of a buffer inside a UMEM chunk on the RX
and TX sides respectively. The \texttt{XDP\_PACKET\_HEADROOM}
reserves the first 256 bytes of each frame for metadata: an
MBUF header (64~B) plus 192~B of headroom that an application
or PMD can use for tunneling headers or pipeline metadata.

Multiple XSKs can share a single UMEM region with their own
per-socket RX/TX/FQ/CQ ring sets. This is what allows a
multi-worker UPF to scale horizontally across NIC RX queues
without copying packets between sockets: a packet steered to
a particular XSK by RSS can be processed and, if needed,
handed to another worker by reference rather than by value.

Unlike DPDK, \texttt{\texttt{AF\_XDP}} does not strictly require hugepages.
Allocating UMEM on hugepages is an optional optimization
that reduces TLB pressure; the kernel will accept any
properly aligned, locked region of memory. The
\texttt{RLIMIT\_MEMLOCK} limit, however, must be high enough
to pin the UMEM, which in production deployments means
raising it to \texttt{unlimited}.

\paragraph{Deployment requirements.}
\texttt{\texttt{AF\_XDP}} sits between \texttt{AF\_PACKET} and DPDK on the deployment
scale: more setup than the kernel stack, considerably less
than full kernel bypass.

\begin{figure}[t]
  \centering
  \includegraphics[width=\columnwidth]{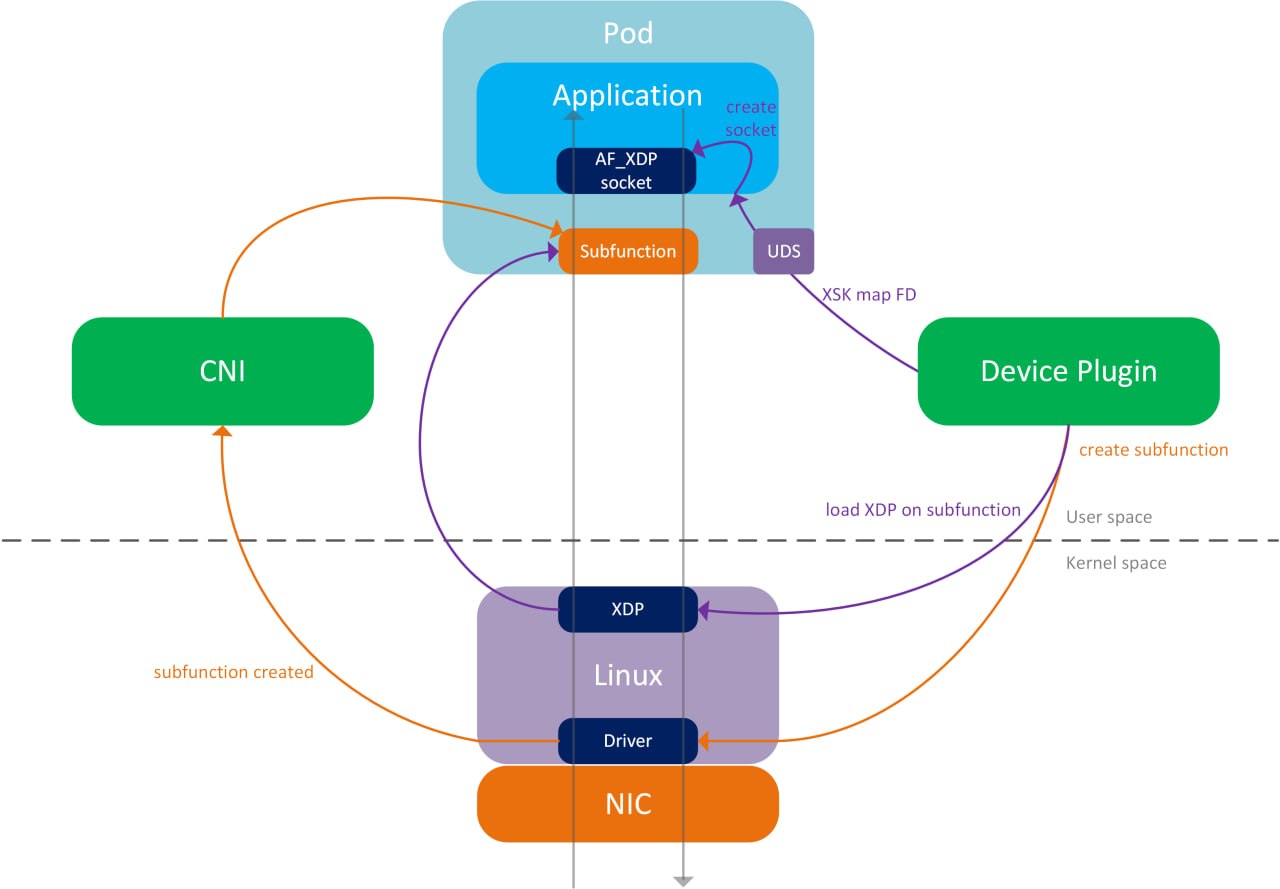}
  \caption{Kubernetes integration of \texttt{\texttt{AF\_XDP}} via the \texttt{\texttt{AF\_XDP}}
    Device Plugin and a CNI. The plugin creates a netdev
    subfunction, loads the XDP program, and passes the XSK
    map file descriptor to the application pod via a Unix
    Domain Socket. The upstream device plugin was archived
    in 2024; our deployment uses the \texttt{host-device}
    CNI pattern instead. \emph{Source: adapted from~\cite{intel-afxdp-guide}.}}
  \label{fig:afxdp-deviceplugin}
\end{figure}

\begin{itemize}[leftmargin=*]
  \item \textbf{NIC driver:} the driver must support XDP at
        the native (\texttt{xdpdrv}) hook, and ideally the
        zero-copy \texttt{\texttt{AF\_XDP}} path. Intel \texttt{i40e} and
        \texttt{ice}, Mellanox \texttt{mlx5}, Broadcom
        \texttt{bnxt}, and several others qualify. The
        \texttt{iavf} driver used by SR-IOV virtual functions
        on Intel hardware does \emph{not} expose the native
        XDP hook, which rules out VF-based \texttt{\texttt{AF\_XDP}}
        deployments on Intel NICs.
  \item \textbf{Kernel:} Linux 5.4 or later, with the
        \texttt{PF\_XDP} socket family compiled in and eBPF
        enabled.
  \item \textbf{Hugepages:} not required, but recommended
        for performance. Our deployment enables 1~GB
        hugepages and configures \texttt{umem.bufcnt} large
        enough to keep the Fill Queue saturated.
  \item \textbf{memlock:} \texttt{RLIMIT\_MEMLOCK} must be
        raised to allow the UMEM region to be locked into
        memory. On a Kubernetes deployment this is
        configured by raising the limit on the underlying
        container runtime (e.g.\ via a systemd
        override for the Canonical Kubernetes \texttt{k8s.service} unit).
  \item \textbf{Privileged container:} required for creating
        XDP sockets and loading the eBPF program. The
        finer-grained capabilities \texttt{CAP\_NET\_RAW},
        \texttt{CAP\_NET\_ADMIN}, \texttt{CAP\_SYS\_ADMIN},
        and \texttt{CAP\_BPF} can substitute for full
        privileged mode in stricter environments.
  \item \textbf{CNI:} the pod must be given a real NIC PF
        (or queue), not a virtual interface that lacks an
        XDP-capable driver. We use the \texttt{host-device}
        CNI to move the physical XXV710 PFs directly into the
        UPF pod; \texttt{macvlan} and similar L2 virtual
        interfaces do not support XDP and cannot be used.
        Figure~\ref{fig:afxdp-deviceplugin} shows the
        canonical Kubernetes wiring through an \texttt{\texttt{AF\_XDP}}
        Device Plugin and CNI: the plugin creates a netdev
        subfunction (or claims a PF queue), loads the XDP
        program, and passes the XSK map file descriptor to
        the application pod through a Unix Domain Socket.
        We note that Intel's upstream \texttt{\texttt{AF\_XDP}} Device Plugin
        repository was archived in
        2024~\cite{afxdp-devplugin-archived}; production
        deployments today increasingly rely on
        \texttt{host-device} CNI directly, or move to CNDP
        for a similar but better-maintained abstraction.
\end{itemize}

\paragraph{Advantages and Limitations.}
\textit{Advantages.}
\texttt{\texttt{AF\_XDP}} delivers most of the throughput benefit of full
kernel bypass without giving up the kernel driver. Zero-copy
and the absence of the kernel network stack make it
substantially faster than \texttt{AF\_PACKET}, while the kernel
remains in charge of the NIC: familiar tools
(\texttt{ethtool}, \texttt{ip}, \texttt{tc}) keep working,
and operators can fall back to the regular stack for
out-of-band traffic by returning \texttt{XDP\_PASS} instead
of redirecting to the XSK. \texttt{\texttt{AF\_XDP}} is also more cloud-native
than DPDK in practice, no hugepages-only constraint, no
\texttt{vfio-pci} binding, and a more conventional security
profile.

\textit{Limitations.}
The XDP-driver requirement is the main constraint. Older
drivers, container-side virtual interfaces, and Intel SR-IOV
VFs (\texttt{iavf}) all lack the native XDP hook.
Performance also depends on the driver's zero-copy support;
copy-mode \texttt{\texttt{AF\_XDP}} (the fallback when the driver lacks ZC) is
still faster than \texttt{AF\_PACKET} but loses a large fraction of
the zero-copy advantage. RSS-based scaling across multiple
XSKs is hampered for GTP-U traffic, because the default RSS
hash sees only the outer IP/UDP headers, not the inner UE
flow; spreading traffic evenly requires either flow director
rules (where supported) or RSS on inner headers through
driver-specific extensions such as Intel ADQ.

\paragraph{Key findings.}
\begin{itemize}[leftmargin=*]
  \item \textbf{Zero-copy is the lever.} The single biggest
        performance step from \texttt{AF\_PACKET} to \texttt{\texttt{AF\_XDP}} comes
        from eliminating the \texttt{sk\_buff} allocation
        and the kernel-to-user copy. On our XXV710/\texttt{i40e} testbed,
        \texttt{ethtool -i} and \texttt{ss --xdp} confirm
        zero-copy is engaged in \texttt{XDP\_DRV} (native
        driver) mode; \texttt{dmesg} reports
        \texttt{MEM\_TYPE\_XSK\_BUFF\_POOL} on the relevant
        RX ring.
  \item \textbf{Throughput.} \texttt{\texttt{AF\_XDP}} sustains 3.69~Mpps
        (1 worker) and 6.47~Mpps (2 workers) full-pipeline at
        64~B, roughly 17$\times$ \texttt{AF\_PACKET} and the highest
        per-core rate of the three kernel-bypass modes,
        reflecting the efficiency of the
        \texttt{net\_\texttt{\texttt{af\_xdp}}} PMD over the zero-copy \texttt{\texttt{AF\_XDP}}
        datapath. RSS-based scaling beyond two workers is
        blocked by the per-netdev XSK socket limit; see
        \S\ref{sec:res-cpu}.

  \item \textbf{Driver support is the gatekeeper.} Our
        deployment only works on physical functions with a
        native-XDP driver (\texttt{i40e} in our case);
        attempts to move the \texttt{\texttt{AF\_XDP}} socket onto an SR-IOV VF
        (\texttt{iavf} driver) fail because that driver does
        not expose the native XDP hook.

  \item \textbf{Memlock and privileged are non-negotiable.}
        Without \texttt{RLIMIT\_MEMLOCK} raised on the host,
        the BESS PMD fails to allocate the UMEM region on
        startup. Without privileged mode (or the right
        capabilities), the XDP program load fails. Both
        conditions are required for the pod to even start.
    \item \textbf{Multi-worker scaling is hampered by RSS.}
        With the in-tree \texttt{i40e} driver's default RSS profile,
        single-flow GTP-U traffic hashes to a single RX
        queue and therefore a single XSK,
        leaving additional worker threads idle. This
        matches the upstream observation in
        \texttt{omec-project/upf} issue~\cite{omec-issue-158} on
        multi-worker scaling caps.
\end{itemize}

\FloatBarrier
\subsection{CNDP}
\label{sec:mode-cndp}

\paragraph{Overview.}
The Cloud Native Data Plane (CNDP)~\cite{cndp-docs} is an
open-source userspace dataplane library originally developed
by Intel and now maintained by the CloudNativeDataPlane
project. CNDP gives applications a DPDK-like programming
model and performance, but layers its userspace stack on
top of \texttt{\texttt{AF\_XDP}} rather than direct \texttt{vfio-pci} kernel
bypass. Packets enter and leave the application through XDP
sockets, just as with plain \texttt{\texttt{AF\_XDP}} (\S\ref{sec:mode-afxdp}),
while the application itself talks to a higher-level CNDP
API that handles logical-port (\texttt{lport}) abstraction,
mempool and MBUF management, threading, and metrics. The
result is a deployment that keeps \texttt{\texttt{AF\_XDP}}'s cloud-native
properties, no hugepages, no VFIO, no PMD binding, kernel
driver retained, while exposing an interface comparable
in expressiveness to DPDK.

In BESS-UPF, CNDP is the default I/O backend in the upstream
\texttt{omec-project/upf} image. A \texttt{CNDPPort} port
type calls the CNDP library directly, rather than going
through a DPDK PMD wrapper, so the integration sits at a
different abstraction layer than the \texttt{AF\_PACKET} and \texttt{\texttt{AF\_XDP}}
modes in Sections~\ref{sec:mode-afpacket}
and~\ref{sec:mode-afxdp}.

\begin{figure*}[t]
  \centering
  \includegraphics[width=0.85\textwidth]{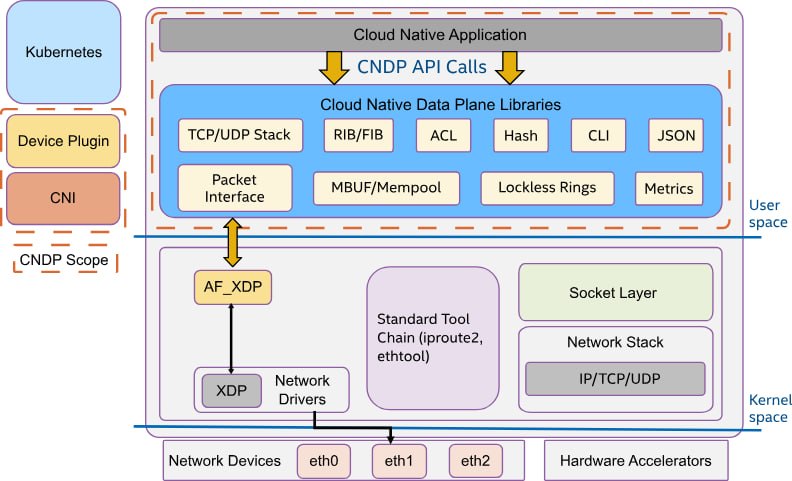}
  \caption{CNDP architecture. Applications call CNDP
    libraries (TCP/UDP stack, RIB/FIB, MBUF/mempool, packet
    interfaces) which sit above the CNDP I/O layer. Packets
    reach the NIC through \texttt{\texttt{AF\_XDP}}, with XDP attached to the
    kernel network drivers. Kubernetes integration uses CNI
    and Device Plugin components. \emph{Source: adapted from~\cite{intel-cndp-guide}.}}
  \label{fig:cndp-arch}
\end{figure*}

\paragraph{Architecture.}
Figure~\ref{fig:cndp-arch} shows the layered CNDP
architecture. The Cloud Native Application sits above the
CNDP API, which is implemented by a set of CNDP libraries
(TCP/UDP stack, RIB/FIB, ACL, hash tables, CLI, JSON,
packet interfaces, MBUF/mempool, lockless rings, metrics).
The CNDP I/O layer below those libraries attaches to the
kernel through \texttt{\texttt{AF\_XDP}} sockets. The kernel side is unchanged
from a plain \texttt{\texttt{AF\_XDP}} deployment: XDP programs in the NIC
driver redirect packets to XSKs through an eBPF XSKMAP.
Kubernetes integration is handled by an optional CNDP
Device Plugin and CNI; deployment can also use a generic
CNI like \texttt{host-device} for simpler setups.

\begin{figure*}[t]
  \centering
  \includegraphics[width=0.85\textwidth]{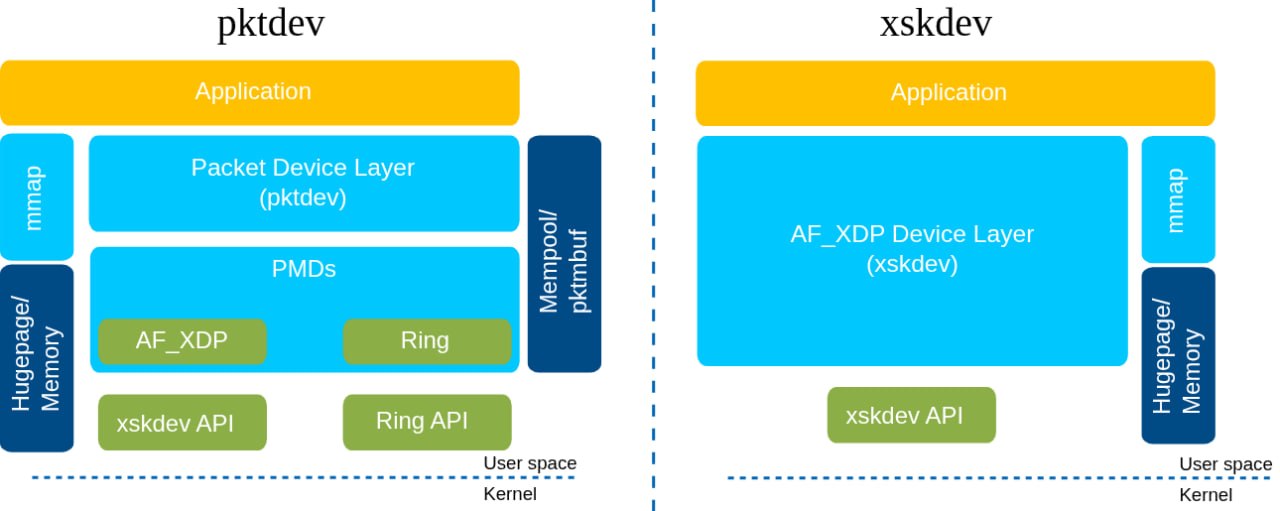}
  \caption{CNDP's two device-layer choices. The
    \texttt{pktdev} layer (left) exposes a PMD-style API
    that can sit on top of \texttt{\texttt{AF\_XDP}}, an internal ring, or
    other transports. The \texttt{xskdev} layer (right) is
    a thinner, \texttt{\texttt{AF\_XDP}}-only API for applications wanting
    direct access. \emph{Source: adapted from~\cite{intel-cndp-guide}.}}
  \label{fig:cndp-pktdev-xskdev}
\end{figure*}

CNDP exposes two device-layer choices, shown in
Figure~\ref{fig:cndp-pktdev-xskdev}. The
\textbf{\texttt{pktdev}} layer is a PMD-style interface
modeled on DPDK's \texttt{rte\_eth\_dev}: applications
request packets via standard receive and transmit calls,
and the PMD chooses which underlying transport (\texttt{\texttt{AF\_XDP}}, an
internal ring, etc.) to use. The \textbf{\texttt{xskdev}}
layer is a thinner, \texttt{\texttt{AF\_XDP}}-only interface for applications
that want direct control. BESS-UPF's \texttt{CNDPPort} uses
the \texttt{pktdev} layer through CNDP's \texttt{\texttt{AF\_XDP}} PMD.

\begin{figure}[t]
  \centering
  \includegraphics[width=\columnwidth]{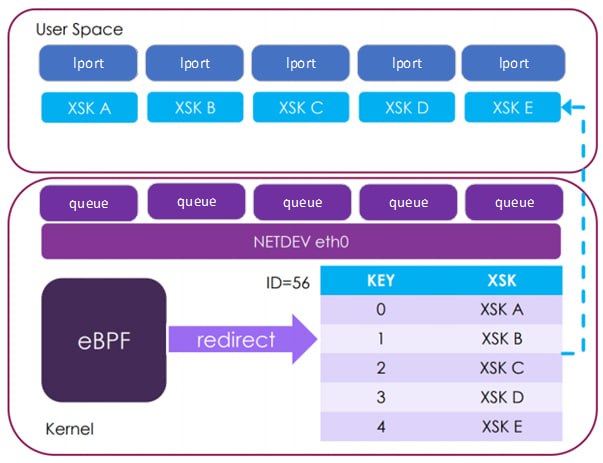}
  \caption{CNDP logical ports (\texttt{lport}) above the
    \texttt{\texttt{AF\_XDP}} sockets (XSK A--E). Each \texttt{lport}
    corresponds to one XSK, which is bound to one NIC RX
    queue. The XDP program in the kernel redirects packets
    to the correct XSK using the queue index as ...the lookup key
    in the XSKMAP. \emph{Source: adapted from~\cite{intel-cndp-guide}.}}
  \label{fig:cndp-lport}
\end{figure}

\paragraph{Datapath packet flow.}
CNDP's datapath mirrors \texttt{\texttt{AF\_XDP}}'s at the kernel level:
incoming packets are intercepted by an XDP program in the
NIC driver and redirected to an XSK via the XSKMAP. CNDP
adds a userspace abstraction on top called the
\textbf{logical port (\texttt{lport})}: a named endpoint
that hides the underlying (\textit{netdev}, queue index)
pair from the application. Figure~\ref{fig:cndp-lport}
shows the relationship: multiple \texttt{lport}s in
userspace map to multiple XSKs in the kernel, each bound
to a NIC RX queue, and the XDP program redirects each
incoming packet to the right XSK using the queue index as
the lookup key.

On receive, a packet follows the same low-level steps as
\texttt{\texttt{AF\_XDP}} (\S\ref{sec:mode-afxdp}, datapath flow), with one
additional layer: the application reads it through the CNDP
\texttt{pktdev} / \texttt{lport} API instead of directly
from the XSK ring. The CNDP library handles descriptor
unpacking, MBUF allocation from a per-port mempool, and
batching. From the application's point of view (BESS-UPF in
our case), packets arrive as MBUFs from a CNDP port, the
\texttt{\texttt{AF\_XDP}} ring mechanics are invisible.

\begin{figure}[t]
  \centering
  \includegraphics[width=\columnwidth]{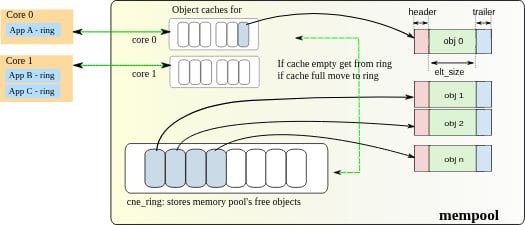}
  \caption{CNDP mempool with per-core object caches backed
    by a central \texttt{cne\_ring}. Each CPU core
    maintains a private cache of MBUFs from the mempool;
    when the cache empties or fills, MBUFs are moved
    between the cache and the central ring. The design
    mirrors DPDK's \texttt{rte\_mempool}
    (see~\S\ref{sec:mode-dpdk}) and avoids cross-core contention on the hot path. \emph{Source: adapted from~\cite{intel-cndp-guide}.}}

  \label{fig:cndp-mempool}
\end{figure}

\paragraph{Memory model.}
CNDP uses the same underlying UMEM mechanism as \texttt{\texttt{AF\_XDP}}
(every packet lives in a UMEM frame, accessible through
mmap'd shared memory) but adds a userspace mempool layer on
top. Figure~\ref{fig:cndp-mempool} shows the per-CPU
mempool design: each core has an object cache for the
mempool's MBUFs, with a central \texttt{cne\_ring} backing
store holding all free objects. The cache absorbs
short-term allocation bursts without contention; the ring
serves as the source of truth when caches drain or
overflow.

The \texttt{umem.bufcnt} parameter in CNDP's JSON
configuration sets the UMEM buffer count. In our deployment
we raise this to 128~K buffers (a 4$\times$ increase over
the upstream default) to keep the Fill Queue saturated
under high-PPS load, a setting we discuss further in
\S\ref{sec:ops}. As with \texttt{\texttt{AF\_XDP}}, hugepages are not
required but are supported as an optimization.

\begin{figure*}[t]
  \centering
  \includegraphics[width=\textwidth]{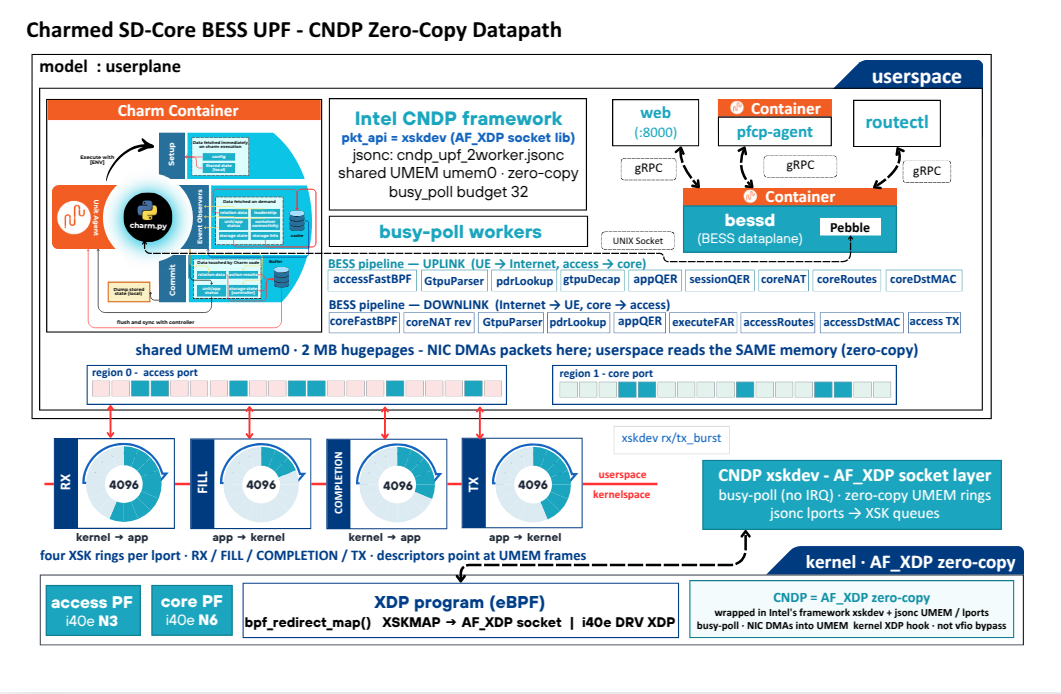}
  \caption{CNDP datapath in the Charmed SD-Core BESS-UPF
    deployment. The CNDP runtime exposes its \texttt{xskdev}
    \texttt{pkt\_api} to the BESS-UPF, configured through
    \texttt{cndp\_upf\_2worker.jsonc}. A shared UMEM region
    (\texttt{umem0}, 2~MB hugepages) is split into access
    and core sub-regions; the NIC DMAs incoming packets
    directly into these UMEM frames and userspace reads the
    same memory without a copy. Four XSK rings per
    \texttt{lport} (RX / Fill / Completion / TX) connect
    busy-poll workers W0/W1 to the kernel-side XDP redirect
    program (\texttt{i40e} driver). The Charm Container on
    the left manages lifecycle.}
  \label{fig:cndp-in-pod}
\end{figure*}

\paragraph{Deployment requirements.}
CNDP's deployment requirements are essentially the union
of \texttt{\texttt{AF\_XDP}}'s requirements with a few CNDP-specific
additions. Figure~\ref{fig:cndp-in-pod} shows the deployed
CNDP datapath in our Charmed BESS-UPF, including the
busy-poll workers and the shared UMEM region split between
access and core ports.

\begin{itemize}[leftmargin=*]
  \item \textbf{NIC driver:} same as \texttt{\texttt{AF\_XDP}}, must
        support native XDP, ideally with zero-copy. Our
        measurements use the \texttt{i40e} driver on
        Intel XXV710 NICs.
  \item \textbf{Kernel:} Linux 5.4+ with \texttt{\texttt{AF\_XDP}} and eBPF
        enabled.
  \item \textbf{CNDP runtime:} the CNDP libraries
        (\texttt{libcndp.so}, the \texttt{cne} libraries,
        and optional Go/Rust bindings) must be installed
        in the container image. The
        \texttt{omec-project/upf} default image already
        ships them.
  \item \textbf{Configuration file:} CNDP is configured by
        a JSON file specifying defaults
        (\texttt{bufcnt}, threads, \texttt{busy\_poll}),
        UMEM regions (\texttt{umems}), and logical ports
        (\texttt{lports}). BESS-UPF's
        \texttt{cndp\_upf\_Nworker.jsonc} files are
        templated by the helm chart.
  \item \textbf{memlock and privileged:} same as \texttt{\texttt{AF\_XDP}},
        with \texttt{RLIMIT\_MEMLOCK} raised and the pod
        running privileged (or with the equivalent
        capability set).
  \item \textbf{CNI:} \texttt{host-device} for simple
        deployments, or a CNDP-aware CNI plus the CNDP
        Device Plugin for multi-port allocation.
  \item \textbf{No hugepages, no \texttt{vfio-pci}, no PMD
        binding:} these are the practical advantages over
        DPDK that motivate CNDP in cloud-native settings.
\end{itemize}

\paragraph{Advantages and Limitations.}
\textit{Advantages.}
CNDP gives applications a DPDK-like programming model
(\texttt{pktdev} ports, MBUF mempools, lockless rings) over
the \texttt{\texttt{AF\_XDP}} datapath. From the application's perspective,
the \texttt{\texttt{AF\_XDP}} ring machinery is hidden behind a clean
\texttt{lport}-based API, and switching between \texttt{\texttt{AF\_XDP}}, an
internal ring, or \texttt{net\_\texttt{af\_packet}} backends is a
configuration change. CNDP is also more cloud-native than
DPDK in operational terms, no hugepages or
\texttt{vfio-pci} required, a smaller container image, and
a more conventional security profile. The
\texttt{omec-project/upf} project has adopted CNDP as its
default backend in recent releases, providing a stable
upstream reference.

\textit{Limitations.}
CNDP inherits \texttt{\texttt{AF\_XDP}}'s constraints, driver-side native
XDP support, no SR-IOV VFs on \texttt{iavf}, RSS hashing
limitations on GTP-U traffic, and adds a few of its own.
The userspace API surface is larger than \texttt{\texttt{AF\_XDP}}'s, so
debugging crosses more layers; fewer Linux distributions
package the CNDP runtime than package \texttt{libxdp}. The
upstream BESS-UPF integration also ships with a
\texttt{coreNAT max\_allowed\_workers=2} annotation; we
patch this out for the present study (raised to 4 for
\texttt{\texttt{AF\_XDP}}/CNDP). The per-netdev \texttt{\texttt{AF\_XDP}} socket limit then
becomes the next ceiling, effectively restricting CNDP to
two workers in practice
(\S\ref{sec:ops}, \S\ref{sec:res-cpu}).

\paragraph{Key findings.}
\begin{itemize}[leftmargin=*]
 \item \textbf{Zero-copy \texttt{\texttt{AF\_XDP}} is engaged.} On our
        XXV710/\texttt{i40e} testbed, \texttt{dmesg} reports
        \texttt{MEM\_TYPE\_XSK\_BUFF\_POOL} on the relevant
        RX ring, and the XDP program is attached in
        driver-native mode (\texttt{xdp\_drv}). The CNDP
        datapath benefits from the full zero-copy \texttt{\texttt{AF\_XDP}}
        path.

 \item \textbf{Full-pipeline ceiling $\approx$5.52~Mpps at
        2 workers.} At 64-byte frames, the BESS-UPF pipeline
        running over CNDP absorbs and forwards 5.52~Mpps
        with zero drops; the per-frame sweep holds flat up
        to 512~B and starts to taper at 1024~B (byte-bound;
        see \S\ref{sec:results}).

    \item \textbf{End-to-end pipeline latency
        $\sim$12.9~$\mu$s mean.} Across 109 million packets
        at 2 workers, the BESS-UPF pipeline adds a P50 of
        13.9~$\mu$s and a P99 of 14.3~$\mu$s end-to-end
        through PDR / QER / FAR lookups and the executeFAR
        action. P99.9 reaches 24.7~$\mu$s, reflecting
        contention on the kernel XDP-redirect path
        introduced by the second worker.
  \item \textbf{Production-stable.} A 24-hour soak at 86~B
        with \texttt{--txonly-multi-flow} delivered 183
        billion packets through the full pipeline with
        zero drops at either \texttt{\texttt{AF\_XDP}} port. Mean throughput
        across 24 hourly samples was 2.100~Mpps with
        3.07\% variance, and \texttt{bessd}'s resident
        memory held constant, no drift, no leak.
    \item \textbf{Multi-worker scaling: stock cap lifted in
        our patched image.} The upstream
        \texttt{omec-project/upf} BESS image carries a
        \texttt{coreNAT max\_allowed\_workers=2} annotation
        that we patch out for this study (raised to 4 for
        \texttt{\texttt{AF\_XDP}}/CNDP, 8 for DPDK; see
        \S\ref{sec:res-cpu}). The per-netdev \texttt{\texttt{AF\_XDP}}
        socket limit then becomes the next ceiling for
        CNDP, restricting it to 2 workers in practice.

\end{itemize}

\FloatBarrier

\subsection{DPDK}
\label{sec:mode-dpdk}

\paragraph{Overview.}
The Data Plane Development Kit (DPDK)~\cite{dpdk} is the
original full-kernel-bypass userspace networking framework, and
remains the reference for raw packet-processing performance on
commodity NICs. Unlike \texttt{\texttt{AF\_XDP}} and CNDP, which keep the kernel
driver in charge of the NIC and intercept packets at the XDP
hook, DPDK takes the NIC away from the kernel entirely: the
device is unbound from its kernel driver, rebound to
\texttt{vfio-pci}, and accessed directly from userspace by a
Poll-Mode Driver (PMD). The kernel network stack is bypassed on
every packet; no \texttt{sk\_buff} is allocated, no socket is
involved, and there are no MSI-X interrupts on the hot path. The
trade-off is the heaviest deployment footprint of the four
modes: hugepages, IOMMU, dedicated CPU cores, and a NIC that
disappears from the kernel's networking view.

In BESS-UPF, DPDK is exposed through a \texttt{PMDPort} using a
native DPDK PMD that matches the underlying NIC: \texttt{net\_i40e}
when the Intel i40e PF is bound to \texttt{vfio-pci},
\texttt{net\_iavf} when an SR-IOV VF on the same family is bound
instead, and \texttt{net\_ice} for E810-class NICs. This is the
same \texttt{PMDPort} integration point used by \texttt{AF\_PACKET} and
\texttt{\texttt{AF\_XDP}} (\S\ref{sec:mode-afpacket}, \S\ref{sec:mode-afxdp}), but
with a native NIC PMD underneath rather than a socket wrapper.

\begin{figure*}[t]
  \centering
  \includegraphics[width=\textwidth]{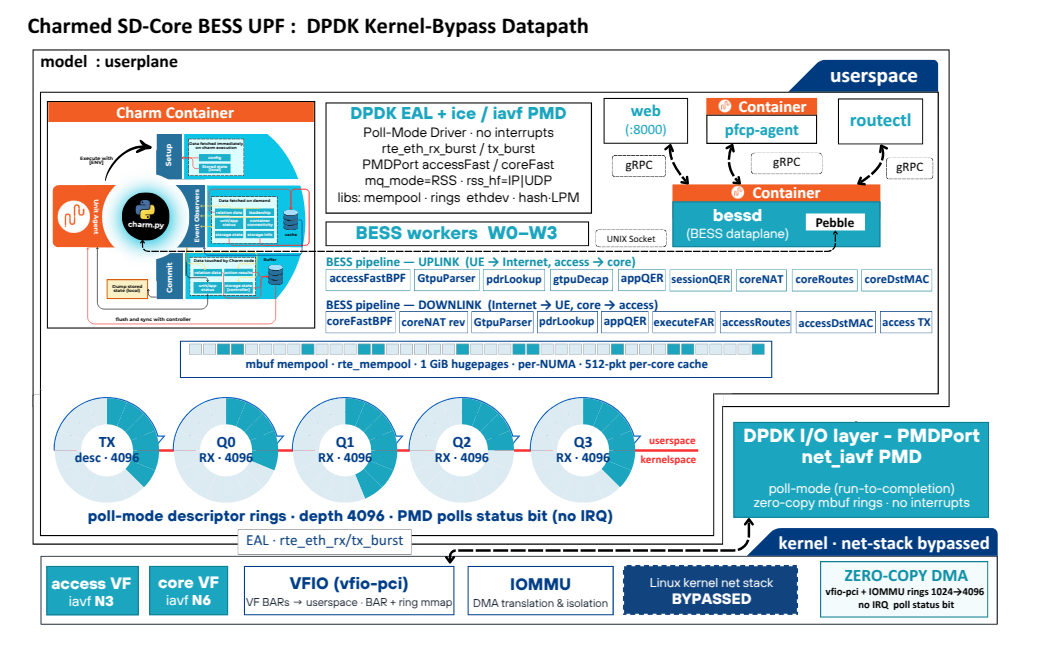}
  \caption{DPDK datapath in the Charmed SD-Core BESS-UPF
    deployment. SR-IOV VFs (\texttt{iavf} N3 and N6) are
    rebound from the kernel to \texttt{vfio-pci}; the DPDK
    PMD polls four 4096-depth descriptor rings (TX + Q0--Q3)
    in run-to-completion across BESS workers W0--W3 pinned
    to CPUs 7--10. The \texttt{rte\_mempool} sits in 1~GiB
    hugepages with a 512-packet per-core cache. The Charm
    Container on the left manages lifecycle and config
    reconciliation; the kernel network stack is fully
    bypassed.}
  \label{fig:dpdk-arch}
\end{figure*}

\paragraph{Architecture.}
Figure~\ref{fig:dpdk-arch} shows the layered DPDK architecture
on the BESS-UPF pipeline. The Environment Abstraction Layer
(EAL) is the foundation: it owns hugepage allocation, NUMA-aware
memory placement, thread pinning, the DPDK device probe, and the
\texttt{ethdev} API that PMDs implement. Above the EAL sit the
data-plane libraries used by the application: \texttt{rte\_mempool}
and \texttt{rte\_mbuf} for packet memory, \texttt{rte\_ring} for
lockless inter-core communication, and helpers for flow tables,
hashes, longest-prefix match, timers, and statistics. The DPDK
I/O layer at the bottom of userspace is the PMD itself, a
poll-mode driver that owns the NIC's RX and TX descriptor rings
and exposes packets to the application as \texttt{rte\_mbuf}
batches.

The kernel side is reduced to a one-time setup role.
\texttt{vfio-pci} handles the initial device handover (Base
Address Register mapping, descriptor-ring \texttt{mmap} into the
DPDK process), and the IOMMU is programmed once to translate and
isolate the device's DMA transactions to the userspace memory
region holding the \texttt{rte\_mbuf} pool. Once the PMD is
initialized, the kernel network stack is bypassed on every
packet: there are no syscalls, no \texttt{sk\_buff} allocations,
no netfilter or traffic-control hooks, and no sockets in the
hot path. The kernel intervenes only on configuration changes
and error paths.

\paragraph{Datapath packet flow.}
On receive, a single packet traverses the following stages:

\begin{enumerate}[leftmargin=*]
  \item The NIC writes the packet into a buffer whose address
        the PMD has already populated in the RX descriptor ring.
        DMA goes directly into a hugepage-backed
        \texttt{rte\_mbuf} sitting in the userspace mempool.
        The IOMMU translates the DMA address to ensure the NIC
        can only reach memory the application owns.
  \item The PMD's poll loop on a dedicated CPU core reads the
        RX ring, observes that the NIC has advanced its
        producer index, and consumes the descriptor. No MSI-X
        interrupt fires; the CPU has been spinning on the ring
        the entire time.
  \item The PMD hands the \texttt{rte\_mbuf} chain to BESS's
        \texttt{PortInc} module. From here the packet flows
        through the BESS module graph (GTP-U decap, PDR
        lookup, QER metering, FAR action) on the same core,
        in a run-to-completion model.
  \item After processing, BESS's \texttt{PortOut} module places
        the \texttt{rte\_mbuf} on the TX descriptor ring of
        the destination port.
  \item The PMD's poll loop writes the NIC's TX doorbell
        register through a memory-mapped I/O write. The NIC's
        on-chip DMA engine reads the descriptor, fetches the
        packet from the hugepage mbuf pool, and transmits it.
  \item Once the NIC has acknowledged completion, the PMD
        reclaims the \texttt{rte\_mbuf} and returns it to the
        mempool's per-core cache for reuse.
\end{enumerate}

Both directions stay entirely in userspace memory accessible
both to the PMD and to the BESS pipeline; the packet never
crosses a kernel-userspace boundary, and there is no copy at
any stage.

\begin{figure}[t]
  \centering
  \includegraphics[width=\columnwidth]{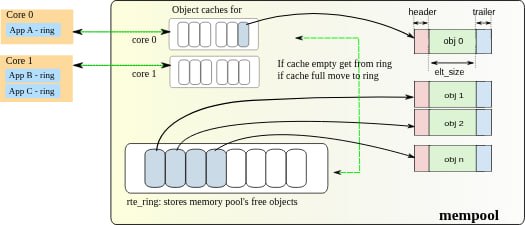}
  \caption{DPDK \texttt{rte\_mempool} with per-core object
    caches backed by a central \texttt{rte\_ring}. Each CPU
    core maintains a private cache of \texttt{rte\_mbuf}s;
    when the cache empties or fills, MBUFs are moved between
    the cache and the central ring. The CNDP mempool
    (\S\ref{sec:mode-cndp}) follows the same design, the
    cache-plus-ring pattern is shared between DPDK and CNDP. \emph{Source: adapted from~\cite{dpdk}.}}
  \label{fig:dpdk-mempool}
\end{figure}

\paragraph{Memory model.}
DPDK's memory model centers on the \texttt{rte\_mempool}, a
hugepage-backed pool of fixed-size \texttt{rte\_mbuf}s used for
both RX and TX packet buffers. Figure~\ref{fig:dpdk-mempool}
shows the per-core cache design: each CPU core that touches the
mempool maintains a private cache of free MBUFs, with a central
\texttt{rte\_ring} as the backing store for all free objects.
When a core's cache empties it bulk-fetches from the ring; when
it fills it bulk-returns to the ring. The cache absorbs most
allocation traffic without touching the contended ring, which
keeps the hot path lock-free.

Memory is allocated from \textbf{hugepages} (typically 2~MB
pages on x86, with 1~GB pages available on supporting CPUs).
Hugepages serve two purposes: they reduce TLB misses on the
large packet-buffer working set (one TLB entry covers a 2~MB
region instead of a 4~KB region), and they provide stable
physical-to-virtual mappings that the IOMMU can program for
NIC DMA without page-migration races. Hugepages are a
\emph{hard requirement} for DPDK, unlike \texttt{\texttt{AF\_XDP}} and CNDP
where they are an optimization.

\paragraph{Deployment requirements.}
DPDK has the most demanding deployment footprint of the four
modes:

\begin{itemize}[leftmargin=*]
  \item \textbf{NIC binding:} the NIC (PF or SR-IOV VF) must
        be unbound from its kernel driver and rebound to
        \texttt{vfio-pci}. The interface disappears from
        \texttt{ip link} and similar kernel-side networking
        tools.
  \item \textbf{IOMMU:} enabled in BIOS and Linux kernel
        (\texttt{intel\_iommu=on iommu=pt} on Intel
        platforms, equivalent options on AMD). Without
        IOMMU, \texttt{vfio-pci} can only run in
        \texttt{noiommu} mode, which is unsafe for
        production deployments.
  \item \textbf{Hugepages:} required. Typically 2~MB or 1~GB
        pages are reserved at boot time (\texttt{hugepages=N}
        kernel parameter) or at runtime via
        \texttt{/sys/kernel/mm/hugepages}. The DPDK
        application maps a configurable amount from this pool
        for its \texttt{rte\_mempool}.
  \item \textbf{CPU isolation:} the PMD's poll loop runs at
        100\% CPU utilization by design. Dedicated cores
        should be removed from the kernel scheduler
        (\texttt{isolcpus} / \texttt{nohz\_full} /
        \texttt{rcu\_nocbs} boot parameters) and pinned to
        the BESS worker thread to avoid scheduler jitter.
  \item \textbf{Privileged container:} required to bind
        \texttt{vfio-pci}, map hugepages, and access
        \texttt{/dev/vfio/}. Specific capabilities
        (\texttt{CAP\_IPC\_LOCK}, \texttt{CAP\_SYS\_ADMIN})
        can substitute for full privileged mode in stricter
        environments.
  \item \textbf{CNI:} the NIC is no longer a kernel netdev,
        so \texttt{host-device} and \texttt{macvlan} CNIs
        do not apply. The SR-IOV Network Device Plugin
        combined with the SR-IOV CNI is the canonical pattern
        for handing a \texttt{vfio-pci}-bound VF to a
        container; for PF-bound deployments,
        \texttt{vfio-pci} device passthrough is configured
        at pod admission time.
  \item \textbf{No kernel networking from the NIC:} the host
        cannot use the NIC for any other traffic, no
        \texttt{ssh} over the same interface, no kernel
        routing, no observability tools that depend on the
        kernel stack. This makes DPDK a poor fit for
        multi-tenant or mixed-purpose hosts.
\end{itemize}

\paragraph{Advantages and Limitations.}
\textit{Advantages.}
DPDK delivers the highest single-host throughput and the
lowest, most deterministic per-packet latency of any of the
four modes. Zero-copy is the default rather than a
driver-dependent optimization, the absence of interrupts
removes scheduler jitter, and the run-to-completion model on
isolated cores keeps the cache hot. DPDK is also the most
mature of the four: it has been in production at line rate
for over a decade, supports a wide range of NICs through
vendor-maintained PMDs, and is the baseline against which all
other userspace networking frameworks are measured.

\textit{Limitations.}
The deployment footprint is the central drawback. Hugepages,
IOMMU, isolated CPU cores, \texttt{vfio-pci} binding, and the
loss of the NIC to the rest of the host all combine to make
DPDK considerably harder to operate in a cloud-native
multi-tenant environment than \texttt{\texttt{AF\_XDP}} or CNDP. The application
also loses access to the kernel's networking features:
\texttt{tc}, \texttt{iptables}, \texttt{netfilter}, eBPF
tracing on the NIC, and standard observability tools all
become unavailable on the bypassed interface. For a single
dedicated UPF on a bare-metal or strongly-isolated host these
costs are usually acceptable; in a Kubernetes cluster sharing
nodes across workloads, the operational overhead frequently
outweighs the performance gain.

\paragraph{Key findings.}
\begin{itemize}[leftmargin=*]
  \item \textbf{Zero-copy by construction.} Unlike \texttt{\texttt{AF\_XDP}},
        where zero-copy is a driver-side optimization that
        may fall back to copy mode, DPDK's datapath is
        zero-copy by design, the NIC DMA target
        \emph{is} the \texttt{rte\_mbuf} the application
        reads. There is no equivalent of \texttt{\texttt{AF\_XDP}}'s copy-mode
        fallback.
    \item \textbf{Throughput.} DPDK sustains 10.30~Mpps
        at 4 workers and 13.09~Mpps at 8 workers
        (pipeline-forwarded, 64~B NDR), plus 16.0~Mpps
        under bidirectional load, the highest of the
        four modes, 1.6--2.5$\times$ ahead of CNDP/\texttt{\texttt{AF\_XDP}}
        at matched worker counts. The native
        \texttt{net\_iavf} PMD over \texttt{vfio-pci}
        avoids the kernel XDP hop and scales past the
        per-netdev \texttt{\texttt{AF\_XDP}} socket limit that caps the other
        two modes; see \S\ref{sec:res-cpu}.

  \item \textbf{Highest CPU efficiency per packet.}
        Eliminating \texttt{sk\_buff} allocation, the full
        kernel stack, and the interrupt path produces the
        lowest cycles-per-packet number of the four modes,
        which translates to the highest packets-per-second
        ceiling on the same hardware.
  \item \textbf{Heaviest deployment cost.} On our testbed,
        bringing up DPDK required pre-boot hugepage
        reservation, kernel command-line changes for IOMMU
        and CPU isolation, container-level privileged
        access, and a separate management NIC for control
        traffic. These steps are not difficult individually
        but they add up to a substantially more
        infrastructure-heavy deployment than the other three
        modes, which can run on a
        stock Canonical Kubernetes cluster with minor configuration tweaks.

  \item \textbf{Best fit for dedicated UPF hosts.} The
        practical sweet spot for DPDK is a UPF deployment on
        a host dedicated to that role, bare metal or
        isolated VM, where the host's NIC is owned by the
        UPF and no other workload competes for it. For
        shared cloud-native deployments, CNDP or \texttt{\texttt{AF\_XDP}}
        give up some of the performance ceiling in exchange
        for substantially better operational ergonomics
        (\S\ref{sec:discuss}).
\end{itemize}

\FloatBarrier

\section{Comparative Analysis}
\label{sec:compare}

Having walked through each of the four modes in
Section~\ref{sec:modes}, we now compare them along the
dimensions that matter most to a UPF operator: how the packet
reaches the application, what hardware and kernel features the
mode requires, what it can reuse from the existing Linux
networking stack, and how much it costs to deploy.
Table~\ref{tab:mode-compare} summarises the architectural
trade-offs side by side, and Figure~\ref{fig:design-space}
places the four modes on a two-axis design space of deployment
complexity versus expected performance ceiling.

\begin{figure*}[t]
  \centering
  \includegraphics[width=\textwidth]{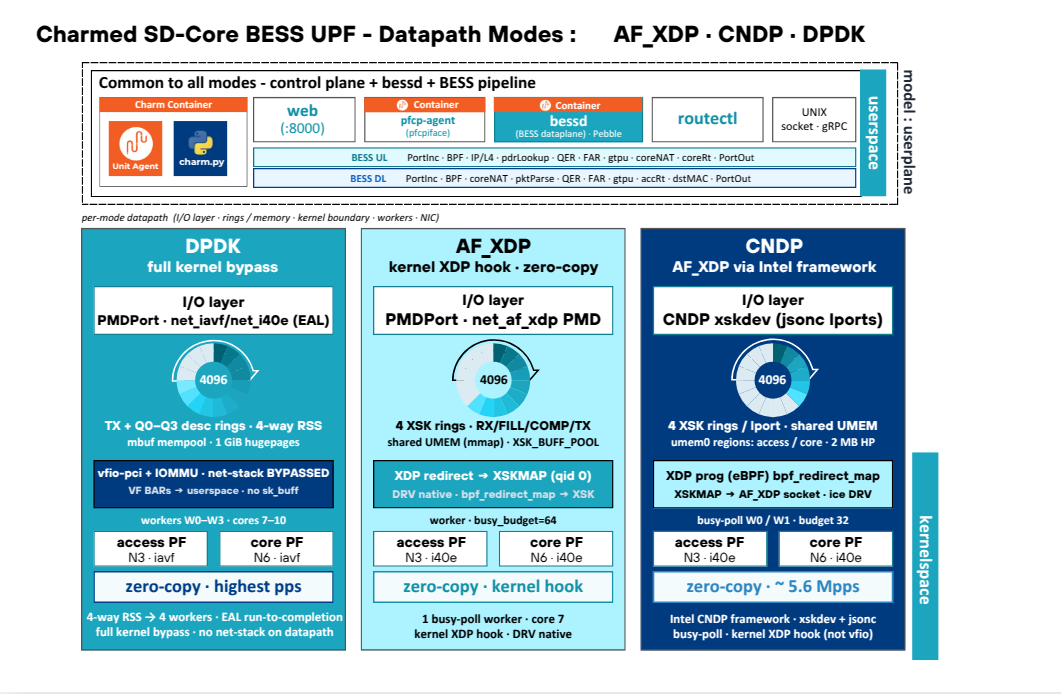}
  \caption{Side-by-side architectural view of the three
    kernel-bypass dataplane modes on the deployed
    Charmed SD-Core BESS-UPF. The top block (Charm
    Container, \texttt{web}, \texttt{pfcp-agent},
    \texttt{bessd}, \texttt{routectl}) and the BESS uplink
    and downlink pipelines are bit-identical across all
    three; only the per-mode I/O layer, ring structure,
    kernel-userspace boundary, and worker pinning change.
    \texttt{AF\_PACKET} is omitted from this view because its
    datapath is qualitatively different (full kernel stack,
    no zero-copy); see Figure~\ref{fig:afpacket-arch}.}
  \label{fig:mode-compare-visual}
\end{figure*}

\begin{table*}[t]
\centering
\caption{Architectural comparison of the four UPF dataplane
  modes across the dimensions that affect performance and
  deployment. Performance rankings are qualitative; absolute
  measured numbers are reported in
  Section~\ref{sec:results}.}
\label{tab:mode-compare}
\small
\renewcommand{\arraystretch}{1.15}
\begin{tabular}{p{3.0cm} p{2.9cm} p{2.9cm} p{2.9cm} p{2.9cm}}
\toprule
\textbf{Dimension} & \textbf{\texttt{\texttt{AF\_PACKET}}} & \textbf{\texttt{\texttt{\texttt{AF\_XDP}}}} & \textbf{CNDP} & \textbf{DPDK} \\
\midrule
\multicolumn{5}{l}{\textit{Datapath properties}} \\
\midrule
Kernel bypass & None & Partial (stack only) & Partial (stack only) & Full \\
Zero-copy & No & Driver-dependent & Driver-dependent (via \texttt{\texttt{AF\_XDP}}) & Yes by design \\
Execution model & Interrupt + NAPI & Configurable (busy-poll) & Configurable (busy-poll) & Poll mode (run-to-completion) \\
Per-packet \texttt{sk\_buff} & Yes & No & No & No \\
Kernel-to-user copy & Yes & No (in ZC mode) & No (in ZC mode) & No \\
\midrule
\multicolumn{5}{l}{\textit{Hardware and kernel requirements}} \\
\midrule
NIC / driver requirement & Any Linux driver & XDP-native driver (no \texttt{iavf} VFs) & XDP-native driver (no \texttt{iavf} VFs) & Driver unbound, rebound to \texttt{vfio-pci} \\
Hugepages & Not required & Optional & Optional & Required \\
IOMMU & Not required & Not required & Not required & Required \\
Dedicated CPU core & Not required & Optional (busy-poll) & Optional (busy-poll) & Required (poll loop $\approx$100\,\%) \\
Privileged container & Not required & Required & Required & Required \\
\midrule
\multicolumn{5}{l}{\textit{Deployment}} \\
\midrule
CNI pattern & Standard (any) & \texttt{host-device} (PF) & \texttt{host-device} or CNDP CNI & SR-IOV CNI / device passthrough \\
BESS port type & \texttt{PMDPort} + \texttt{net\_\texttt{af\_packet}} & \texttt{PMDPort} + \texttt{net\_\texttt{\texttt{af\_xdp}}} & \texttt{CNDPPort} & \texttt{PMDPort} + native PMD (\texttt{net\_i40e}, \texttt{net\_ice}, \texttt{net\_iavf}) \\
NIC visible to kernel & Yes & Yes & Yes & No (rebound to \texttt{vfio-pci}) \\
Kernel feature reuse (\texttt{tc}, netfilter, eBPF) & Full & Partial (\texttt{XDP\_PASS} fallback) & Partial (\texttt{XDP\_PASS} fallback) & None \\
\midrule
\multicolumn{5}{l}{\textit{Operational properties (qualitative)}} \\
\midrule
Deployment complexity & Low & Medium & Medium & High \\
Cloud-native fit & High & High & High & Low \\
Throughput ranking & Lowest & Mid & Mid & Highest \\
Latency ranking & Highest & Mid & Mid & Lowest \\
\bottomrule
\end{tabular}
\end{table*}

\begin{figure*}[t]
\centering
\includegraphics[width=0.7\textwidth]{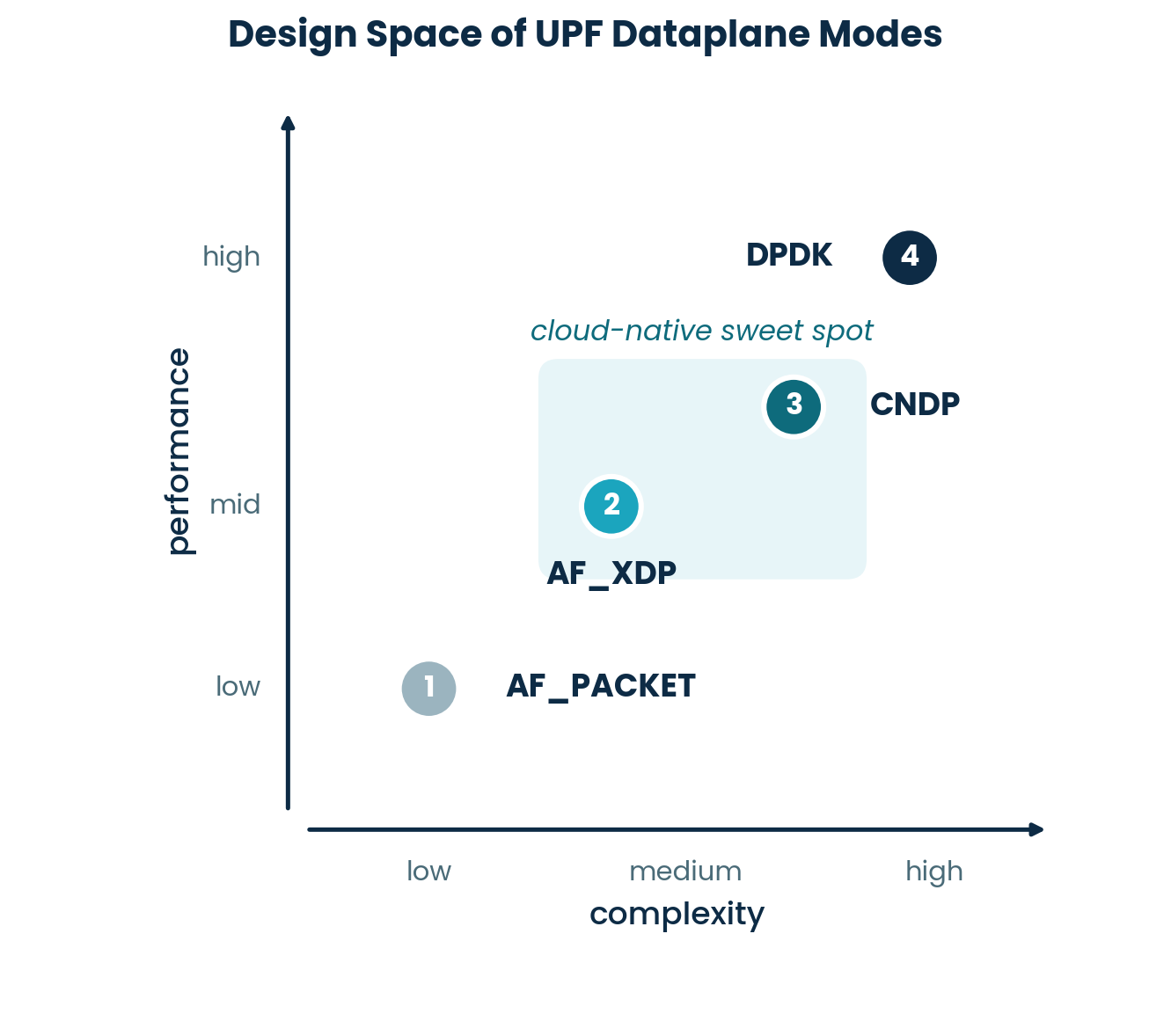}
\caption{Design space of UPF dataplane modes. \texttt{AF\_PACKET} sits
  at low complexity and low performance; DPDK at high
  complexity and high performance; \texttt{\texttt{AF\_XDP}} and CNDP cluster
  in the cloud-native middle, where stock-stack throughput is
  available without DPDK's hugepage, VFIO, and isolated-core
  requirements.}
\label{fig:design-space}
\end{figure*}

\paragraph{The spectrum.}
Figure~\ref{fig:mode-compare-visual} and
Table~\ref{tab:mode-compare} present the comparison
visually and dimensionally. The four modes form a clear
spectrum of \emph{kernel involvement}. \texttt{\texttt{AF\_PACKET}} keeps the kernel network stack on the
hot path; \texttt{\texttt{\texttt{AF\_XDP}}} bypasses the stack but keeps the driver;
CNDP wraps \texttt{\texttt{\texttt{AF\_XDP}}} with a userspace abstraction that brings
DPDK-like programming ergonomics; and DPDK takes the NIC away
from the kernel entirely. Performance moves in lockstep with
this progression, each step away from the kernel removes a
class of per-packet overhead (\texttt{sk\_buff} allocation,
stack traversal, copy-to-user, interrupt handling) but adds a
class of deployment complexity (XDP-capable driver, hugepages,
\texttt{vfio-pci} binding, isolated cores).

\paragraph{The cloud-native gap.}
The axis that matters most to a Kubernetes operator is not
raw performance but \emph{deployment ergonomics under
cloud-native constraints}. By this measure, \texttt{\texttt{AF\_XDP}} and CNDP
occupy a clear sweet spot: they deliver close to DPDK's
throughput with hugepages optional, no \texttt{vfio-pci}
binding, and the NIC retained under the kernel for standard
diagnostic tools. DPDK delivers higher throughput but at the
cost of a substantially heavier operational footprint that is
hard to justify on shared nodes. \texttt{AF\_PACKET} delivers the
lowest throughput but is the only mode that runs unmodified
on any Kubernetes cluster, which makes it the right starting
point for development and the wrong choice for production at
scale.

\paragraph{The single experimental variable.}
For our measurements (Section~\ref{sec:results}), the design
shown in Figure~\ref{fig:bess-pipeline} ensures that the only
variable between runs is the BESS port driver. The BESS
module graph, the PFCP/PDR tables, the GTP-U logic, and the
metering rules remain bit-identical across all four
configurations. Any performance difference reported in
Section~\ref{sec:results} is therefore attributable to the
I/O backend, not to differences in the pipeline.

\section{Experimental Methodology}
\label{sec:method}

This section describes the testbed, metrics, traffic
generation, and reproducibility setup used to evaluate the
four UPF dataplane modes. Every measurement reported in this
paper (throughput sweeps, latency distributions,
multi-worker scaling, the 24-hour soak, and the real-RAN
end-to-end run) was produced by \benchname{},
 our UPF benchmarking framework. Measurements were
collected on a single physical testbed
(\S\ref{sec:method-testbeds}). The control
argument (\S\ref{sec:method-control}) explains what is held
constant across all runs; \S\ref{sec:method-metrics} defines
the metrics; \S\ref{sec:method-traffic} describes the load
generators and traffic profiles; and \S\ref{sec:method-repro}
lists the configuration files, scripts, and helm values used
so the results can be reproduced.

\subsection{Testbed}
\label{sec:method-testbeds}

\begin{table*}[t]
\centering
\caption{Testbed configuration. All four dataplane modes were
  evaluated on the same host, same NIC, same kernel, same
  Kubernetes cluster, and the same BESS-UPF image. The only
  variable across runs is the BESS port driver.}
\label{tab:testbeds}
\small
\renewcommand{\arraystretch}{1.15}
\begin{tabular}{p{4cm} p{10cm}}
\toprule
\textbf{Component} & \textbf{Configuration} \\
\midrule
Platform & x86\_64 (Intel 64), little-endian, SSE4.2 / AVX2 baseline \\
Host & \texttt{three} (\texttt{10.8.0.4}), 2$\times$ Intel Xeon Gold 6148 @ 2.40~GHz, 40 cores total (20 cores/socket, 2 NUMA nodes, HT off) \\
Memory & 125~GiB; 24 $\times$ 1~GiB hugepages on NUMA node 0 \\
Kernel & 5.15.0-1079-realtime (PREEMPT\_RT), \texttt{mitigations=off}, \texttt{intel\_pstate=disable}, governor \texttt{performance}, \texttt{idle=poll} \\
NIC & Intel XXV710 25~GbE (\texttt{i40e}), PCI 0000:18:00.x, on-chip VEB for hairpin \\
SR-IOV VFs & 18:0a.0--.5 (access/core PFs + generator VFs, all \texttt{vfio-pci} for DPDK; access/core PFs for \texttt{\texttt{AF\_XDP}}/CNDP) \\
CPU isolation & \texttt{isolcpus=domain,0-15} + \texttt{isolcpus=0-25}, \texttt{rcu\_nocbs=0-15}, \texttt{kthread\_cpus=16-39}, \texttt{nohz/skew\_tick} \\
\texttt{bessd} worker pins & cores 5--6 (1--2w CNDP/\texttt{\texttt{AF\_XDP}}); cores 7--10 (4w DPDK); generator on cores 8--11 \\
IOMMU & \texttt{intel\_iommu=on iommu=pt} \\
Kubernetes & Canonical Kubernetes (k8s) + charmed sd-core / aether-onramp~\cite{aether-onramp} \\

CNI & \texttt{host-device} (PF or VF in pod) \\
BESS-UPF image & \texttt{omec-project/upf-bess} (with mode-specific patches as documented in \S\ref{sec:ops}) \\
\bottomrule
\end{tabular}

\end{table*}

\begin{figure*}[t]
  \centering
  \includegraphics[width=0.85\textwidth]{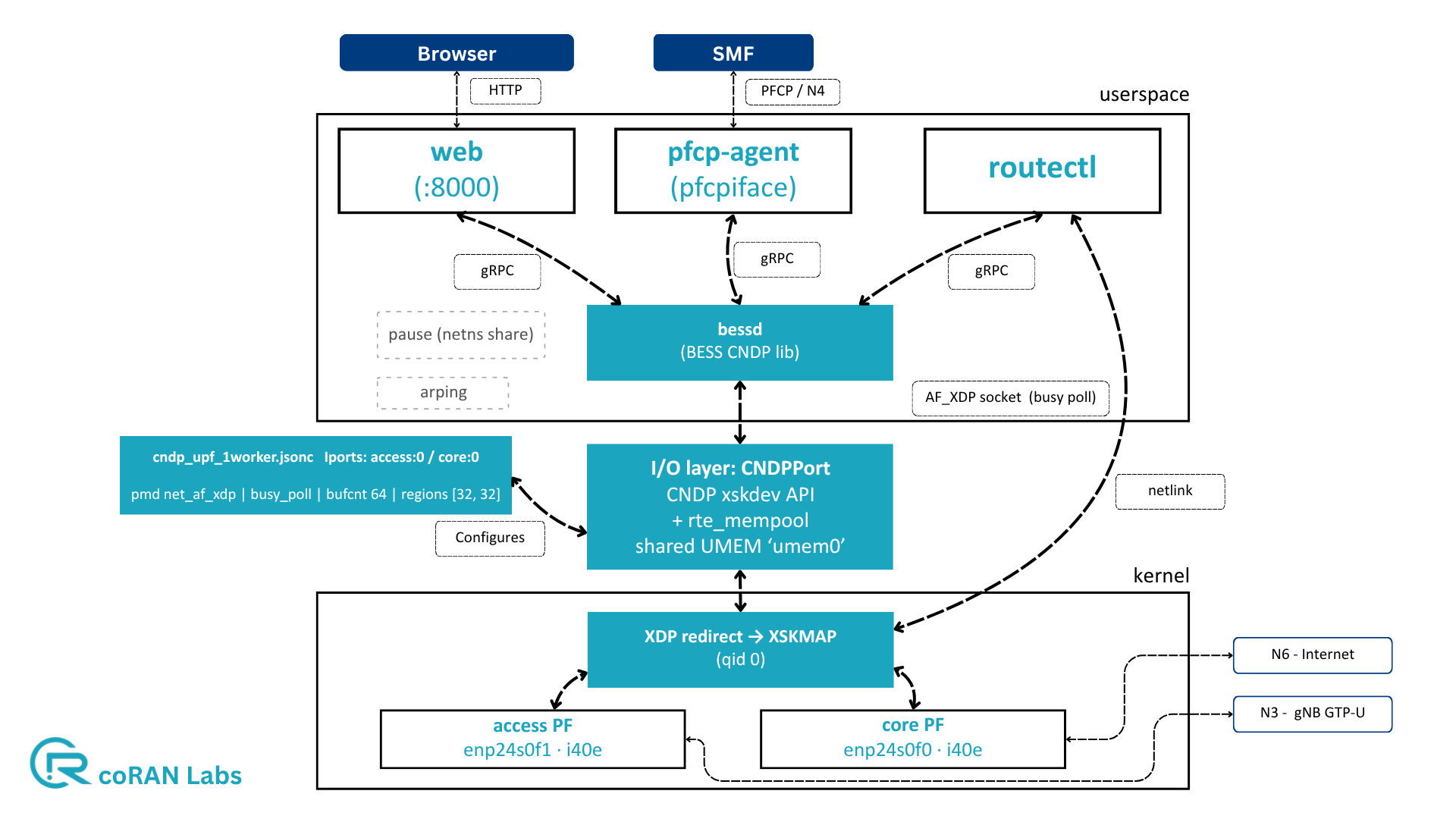}
 \caption{Testbed deployment with CNDP as the I/O backend.
    The same pod composition, helm chart, and PFCP control
    plane are used for all four dataplane modes; only the
    BESS port type (\texttt{CNDPPort} here, \texttt{PMDPort}
    + a PMD for the other three) and the NIC binding change
    between runs.}

  \label{fig:tb-cndp}
\end{figure*}

\paragraph{Testbed.}
All measurements were collected on a single host,
\texttt{server three} (\texttt{10.8.0.4}): a dual-socket
2$\times$Xeon Gold 6148 with 24 reserved 1~GiB hugepages on
NUMA node 0 and a PREEMPT\_RT kernel
(\texttt{5.15.0-1079-realtime}). The NIC is an Intel
XXV710 25~GbE (\texttt{i40e}); its access and core PFs act
as the N3 and N6 interfaces, with SR-IOV VFs hosting both
the UPF datapath ports and the synthetic generators. The
\texttt{i40e} on-chip VEB hairpins generator-VF traffic
directly into the UPF access PF without crossing a physical
wire, allowing the UPF's internal processing capacity to be
measured independently of the lab's 1~GbE optics. CPU isolation is configured at boot
(\texttt{isolcpus=domain,0-15}, \texttt{rcu\_nocbs=0-15},
\texttt{kthread\_cpus=16-39}); the \texttt{bessd} worker
threads are pinned to cores 5--6 for the CNDP and \texttt{\texttt{AF\_XDP}}
runs (1--2 workers) and to cores 7--10 for the DPDK runs
(up to 4 workers).
Figure~\ref{fig:tb-cndp} shows the deployed pod composition
when CNDP is the I/O backend; the same composition is used for
the other three modes with only the BESS port driver swapped.
The software stack (charmed sd-core / aether-onramp BESS-UPF, the
helm chart, the PFCP control plane, and the BESS module graph)
is identical across all four configurations. The only variable between runs is the BESS port driver and
its underlying PMD or library.

\paragraph{Deployment model.}
The UPF is deployed as a Charmed operator within the
upstream \texttt{charmed sd-core} bundle. A Juju controller
manages the lifecycle of each Charmed unit (UPF, AMF, SMF,
PCF, UDR, NRF, UDM, NSSF, AUSF) through a Unit Agent and a
Python charm (\texttt{charm.py}) running inside the pod's
charm sidecar container. Configuration is declarative: the
BESS port driver, the CNDP \texttt{jsonc} parameters,
worker counts, and CPU pinning are set as charm config
options that are reconciled through the Juju
observe--analyse--act loop without re-rendering manifests by
hand. This matters for the experiment in two ways. First,
switching the dataplane between \texttt{AF\_PACKET}, \texttt{\texttt{AF\_XDP}}, CNDP,
and DPDK is a one-line charm config change rather than a
helm-chart edit, which keeps the everything-else-constant
guarantee easy to enforce. Second, the charm-mediated
deployment is closer to what an operator would actually run
in production than a hand-deployed pod, so the soak and
real-RAN results in Section~\ref{sec:results} carry over
without translation. Figure~\ref{fig:tb-cndp} shows the
deployed pod composition, including the charm sidecar and
the Juju controller relation.

\subsection{Held Constant vs.\ Varied}
\label{sec:method-control}

The central design choice of this study is to hold the
entire UPF \emph{pipeline} constant and vary only the BESS
\emph{port driver}.

\begin{itemize}[leftmargin=*]
  \item \textbf{Held constant across all runs:} the BESS
        module graph (\texttt{PortInc} $\rightarrow$
        \texttt{GtpuDecap} $\rightarrow$ \texttt{WildcardMatch}
        (PDR) $\rightarrow$ QER metering $\rightarrow$ FAR
        action $\rightarrow$ \texttt{PortOut}); the PFCP
        control plane; the PDR/FAR/QER rules installed via
        \texttt{pybess}; the metering thresholds; the helm
        chart used to template the pod; the CNI used to give
        the pod its PFs; and the BESS userspace code itself.
  \item \textbf{Varied per run:} the BESS port type
        (\texttt{PMDPort} + DPDK PMD vs.\ \texttt{CNDPPort}),
        the underlying PMD or library
        (\texttt{net\_\texttt{af\_packet}} / \texttt{net\_\texttt{\texttt{af\_xdp}}} /
        \texttt{net\_i40e} / CNDP runtime), and the NIC
        binding (kernel driver vs.\ \texttt{vfio-pci}).
\end{itemize}

Any performance difference we report in
Section~\ref{sec:results} is therefore attributable to the
I/O backend, not to differences in the pipeline.

One stimulus-side caveat applies to this attribution: the
synthetic generator is not identical across modes
(\texttt{testpmd} for the DPDK runs, kernel \texttt{pktgen}
for \texttt{AF\_XDP} and CNDP, for the RSS reason explained
in \S\ref{sec:method-traffic}). The UPF pipeline, NIC,
kernel, and worker placement are held constant, so the I/O
backend remains the controlled variable, but the generator
difference is a known limit on the strict single-variable
framing and is detailed in \S\ref{sec:method-traffic}.

\subsection{Metrics}
\label{sec:method-metrics}

We report four classes of metrics.

\paragraph{Throughput.}
Packets per second (Mpps) and bits per second (Gbps)
sustained through the full BESS-UPF pipeline at the egress
port, swept across frame sizes (64, 128, 256, 512, 1024,
1518~B per RFC~2544). We distinguish \emph{absorbed} (packets received by
the ingress BESS port after the \texttt{\texttt{AF\_XDP}} / DPDK ring) from
\emph{forwarded} (packets that successfully exit the egress
port after traversing the pipeline). The headline figure is
\emph{forwarded throughput}.

\paragraph{Latency and jitter.}
End-to-end pipeline latency, measured by inserting BESS
\texttt{Timestamp} and \texttt{Measure} module pairs at
\texttt{pktParse:1} (ingress, after parse) and
\texttt{executeFAR:1} (egress, after the FAR action) via the
\texttt{pybess} gRPC interface. The measurement captures the
time a packet spends in the entire stateful chain (PDR
lookup, QER metering, FAR lookup, FAR action). We report
mean, median (P50), P90, P95, P99, P99.9, and P99.99
percentiles, plus jitter at each percentile (defined as the
deviation from the P50). All values are in nanoseconds with
100~ns measurement resolution.

\paragraph{CPU and core scaling.}
Per run we record \texttt{bessd} worker CPU utilization, the
number of active BESS workers, the per-worker packet rate,
and the multi-worker scaling efficiency observed when we
attempt to add workers. The stock 
upstream \texttt{coreNAT max\_allowed\_workers=2}
annotation was patched out for this study (raised to 4 for
\texttt{\texttt{AF\_XDP}}/CNDP and 8 for DPDK; see \S\ref{sec:res-cpu}).

\paragraph{Loss and stability.}
Drops are tracked at both BESS ports (ingress and egress) and
reconciled against the generator's offered count. Stability
is verified by 24-hour sustained-load soaks at saturation,
recording cumulative RX/TX counters and \texttt{bessd}
resident memory hourly.

\subsection{Traffic and Tools}
\label{sec:method-traffic}

We use both synthetic packet generators (for raw I/O
ceilings) and real RAN-side traffic (for end-to-end
measurements).

\paragraph{Synthetic load.}
Two generators are used. For DPDK, we use \texttt{testpmd}
in \texttt{txonly} mode with \texttt{--txonly-multi-flow} on
a \texttt{vfio-pci}-bound VF, which works correctly with the
native \texttt{net\_iavf} PMD's RSS profile. For \texttt{\texttt{AF\_XDP}} and
CNDP, however, \texttt{testpmd --txonly-multi-flow} varies the
wrong field, the resulting packets all hash to a single
RX queue (worker~1 idle, no measurable 2-worker scaling). We
therefore use kernel \texttt{pktgen} with \texttt{IPDST\_RND}
(destination IP randomised), which delivers a clean 50/50
queue split and the expected near-linear 1$\to$2 worker scaling.
Both generators hairpin through the on-chip i40e VEB into
the UPF access PF and never cross a physical wire, isolating
the UPF's internal processing capacity from the lab's 1~GbE
optics. Frame sizes are swept across
64, 128, 256, 512, 1024, and 1518~B per RFC~2544. The
24-hour soak run is reported separately at the 86~B
saturation point.

\paragraph{Stateful TCP/UDP load.}
\texttt{iperf3} provides TCP and UDP load over the same
pipeline, both for synthetic single-flow tests and for
end-to-end tests where the traffic originates from a UE
attached over the RAN.

\paragraph{End-to-end latency.}
\texttt{ping} (ICMP) and \texttt{speedtest-cli} (HTTPS)
provide end-to-end round-trip-time and TCP-throughput
measurements with the UE attached through the RAN,
exercising the full path from UE through gNB and UPF to the
data network.

\paragraph{RAN source.}
End-to-end measurements use two different RAN sources of
increasing realism. The default driver is OAI
\texttt{rfsim} acting as a software gNB on the N3 link,
useful for control-plane plumbing and PFCP-path validation
but not for throughput ceilings. For functional validation
under real-radio conditions we additionally run a
\emph{disaggregated 5G RAN}: an OCUDU 7.2-split CU/DU pair
(O-RAN-aligned, fronthaul over eCPRI) terminating in a
Liteon Open RAN radio unit (RU), with a commercial UE
attaching over the air. Registration, authentication, PDU
session establishment, and bearer setup are all driven by
the real UE through this disaggregated chain into the
Charmed SD-Core control plane, with user-plane traffic
landing on the deployed UPF unmodified. We note explicitly
that both RAN sources cap the achievable RAN-to-UPF
throughput well below what the testbed can sustain on the
I/O side; for the I/O-ceiling measurements in
Section~\ref{sec:results}, the synthetic \texttt{testpmd}
generator is therefore the authoritative source. The
disaggregated setup serves as functional ground truth
(``does the deployed UPF actually carry real subscriber
traffic end-to-end?''), not as a throughput benchmark.

\subsection{Reproducibility}
\label{sec:method-repro}

The \texttt{pybess} scripts that install the PDR/FAR/QER
rules and insert the \texttt{Timestamp}/\texttt{Measure}
latency probes, together with the \texttt{testpmd},
\texttt{iperf3}, and \texttt{ping} invocations, are
described in this section so that the campaign can be
reproduced on an equivalent testbed.

\FloatBarrier

\section{Results and Evaluation}
\label{sec:results}

This section reports the measurements collected on the
testbed described in Section~\ref{sec:method}. We organise
the results around the four metric classes defined in
\S\ref{sec:method-metrics}: throughput
(\S\ref{sec:res-throughput}), latency and jitter
(\S\ref{sec:res-latency}), and CPU and core scaling
(\S\ref{sec:res-cpu}) are reported as cross-mode comparisons;
stability and production readiness (\S\ref{sec:res-stability})
are validated in depth on CNDP, the mode used in our
production deployment. Throughput-ceiling and stability
numbers are obtained with the synthetic \texttt{testpmd}
generator at line rate; latency is measured in-pipeline via
the BESS \texttt{Timestamp}/\texttt{Measure} probe pair.

\subsection{Throughput}
\label{sec:res-throughput}

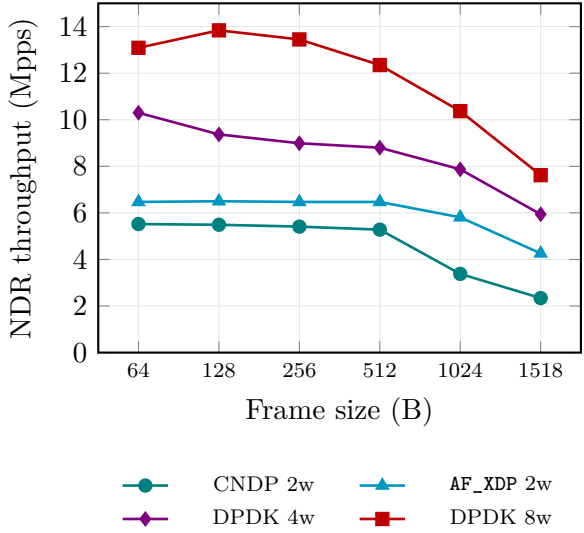
\begin{figure}[t]
\centering
\begin{tikzpicture}
\begin{axis}[
  width=\columnwidth, height=6.2cm,
  xlabel={Frame size (B)},
  ylabel={NDR throughput (Mpps)},
  symbolic x coords={64,128,256,512,1024,1518},
  xtick=data,
  x tick label style={font=\scriptsize},
  ymin=0, ymax=15,
  ytick={0,2,4,6,8,10,12,14},
  grid=both,
  grid style={gray!20},
  mark size=2.2pt,
  thick,
  legend style={
    at={(0.5,-0.32)},
    anchor=north,
    legend columns=2,
    font=\scriptsize,
    draw=none,
    column sep=0.5cm,
    row sep=0.05cm,
  },
]
\addplot[color=teal, mark=*, line width=1pt] coordinates {
  (64,5.52) (128,5.49) (256,5.41) (512,5.28) (1024,3.38) (1518,2.34)
};
\addlegendentry{CNDP 2w}
\addplot[color=cyan!75!black, mark=triangle*, line width=1pt] coordinates {
  (64,6.47) (128,6.50) (256,6.47) (512,6.47) (1024,5.81) (1518,4.26)
};
\addlegendentry{\texttt{\texttt{AF\_XDP}} 2w}
\addplot[color=violet, mark=diamond*, line width=1pt] coordinates {
  (64,10.30) (128,9.37) (256,8.99) (512,8.80) (1024,7.87) (1518,5.94)
};
\addlegendentry{DPDK 4w}
\addplot[color=red!75!black, mark=square*, line width=1pt] coordinates {
  (64,13.09) (128,13.84) (256,13.45) (512,12.35) (1024,10.37) (1518,7.62)
};
\addlegendentry{DPDK 8w}
\end{axis}
\end{tikzpicture}

\caption{NDR (Non-Drop Rate) throughput vs.\ frame size on
  the XXV710/\texttt{i40e} testbed. DPDK 8w leads through
  all frame sizes (13.1~Mpps at 64~B); DPDK 4w and \texttt{\texttt{AF\_XDP}} 2w
  sit between $\approx$6.5 and 10.3~Mpps at small frames;
  CNDP 2w sits just below at 5.5~Mpps. Large frames
  (1024~B+) become byte-bound at $\approx$28--92~Gbps
  internal VEB capacity. \texttt{AF\_PACKET} ($\approx$0.25~Mpps, flat across all frame sizes)
  is omitted from this chart for legibility; see
  Table~\ref{tab:throughput-cross}.
  }

\label{fig:throughput-sweep}
\end{figure}

Figure~\ref{fig:throughput-sweep} plots forwarded
full-pipeline throughput against frame size for all four
modes, and Table~\ref{tab:throughput-cross} summarises the
headline rate at the 64~B saturation point.
Table~\ref{tab:throughput-cndp} gives the detailed per-frame
sweep for all three kernel-bypass modes.

\begin{table*}[t]
\centering
\caption{Cross-mode NDR throughput at 64~B (per RFC~2544 /
  ETSI NFV-TST~009) on the XXV710/\texttt{i40e} testbed.
  Fwd/Abs~=~1.000 (zero pipeline loss) across all reported
  configurations. \texttt{AF\_PACKET} is reported at 86~B as a
  kernel-stack reference point; the kernel-bypass modes at
  64~B.
}
\label{tab:throughput-cross}
\small
\renewcommand{\arraystretch}{1.2}
\begin{tabular}{l r r r r l}
\toprule
\textbf{Mode} & \textbf{1w} & \textbf{2w} & \textbf{4w} & \textbf{8w}$^{\ddagger}$ & \textbf{Notes} \\
\midrule
\texttt{AF\_PACKET}$^{\dagger}$ & 0.224 & 0.252 &, &, & 86~B reference \\
CNDP (xskdev)          & 2.65  & 5.52  &, &, & 64~B, 2.09$\times$ scaling \\
\texttt{\texttt{AF\_XDP}} (net\_\texttt{\texttt{af\_xdp}}) & 3.69  & 6.47  &, &, & 64~B, 1.75$\times$ scaling \\
DPDK (net\_iavf)       & 2.99  & 5.56  & 10.30 & 13.09 & 64~B, 3.44$\times$ scaling 1w$\to$4w \\
DPDK bidirectional     &,   &,   & 10.28 & 16.0  & 64~B, UL+DL on same UPF \\
\bottomrule
\end{tabular}

\vspace{0.3em}
{\footnotesize $^{\dagger}$AF\_PACKET is reported as a
qualitative kernel-stack reference; the order-of-magnitude
gap to the kernel-bypass modes is structural to the
AF\_PACKET datapath (full kernel stack + copy-to-user).}

\vspace{0.2em}
{\footnotesize $^{\ddagger}$DPDK 8w numbers are obtained
under VEB recirculation oversubscription, since the
\texttt{testpmd} generator is CPU-bound at
$\approx$10.5~Mpps
(\S\ref{sec:method-traffic}, \S\ref{sec:future}). The
1w--4w points lie within the generator's clean
unidirectional range; the 8w points should be read as
internal-capacity figures, not directly generated NDR
points.}
\end{table*}

\begin{table*}[t]
\centering
\caption{Per-mode NDR throughput sweep on XXV710/\texttt{i40e},
  64--1518~B, NDR at 0\,\% frame loss, 10~s snapshot-delta
  per point. CNDP and \texttt{\texttt{AF\_XDP}} at 2 workers; DPDK at 4 and 8
  workers.}
\label{tab:throughput-cndp}
\small
\renewcommand{\arraystretch}{1.15}
\begin{tabular}{r r r r r r}
\toprule
\textbf{Frame (B)} & \textbf{CNDP 2w} & \textbf{\texttt{AF\_XDP} 2w} & \textbf{DPDK 4w} & \textbf{DPDK 8w}$^{\ddagger}$ & \textbf{Limit} \\

\midrule
64   & 5.52 & 6.47 & 10.30 & 13.09 & CPU/packet \\
128  & 5.49 & 6.50 &  9.37 & 13.84 & CPU/packet \\
256  & 5.41 & 6.47 &  8.99 & 13.45 & CPU/packet \\
512  & 5.28 & 6.47 &  8.80 & 12.35 & mixed \\
1024 & 3.38 & 5.81 &  7.87 & 10.37 & byte/bandwidth \\
1518 & 2.34 & 4.26 &  5.94 &  7.62 & byte/bandwidth \\
\bottomrule
\end{tabular}

\vspace{0.3em}
{\footnotesize $^{\ddagger}$DPDK 8w numbers are obtained
under VEB recirculation oversubscription, since the
\texttt{testpmd} generator is CPU-bound at
$\approx$10.5~Mpps
(\S\ref{sec:method-traffic}, \S\ref{sec:future}). The
1w--4w points lie within the generator's clean
unidirectional range; the 8w points should be read as
internal-capacity figures, not directly generated NDR
points.}
\end{table*}

The headline numbers are: \texttt{AF\_PACKET} 0.252~Mpps, CNDP 5.52~Mpps
(2w), \texttt{\texttt{AF\_XDP}} 6.47~Mpps (2w), and DPDK 13.09~Mpps (8w),
all at 64~B NDR. Three observations stand out. First,
\texttt{AF\_PACKET} sits an order of magnitude below the kernel-bypass
modes, as expected from its kernel-stack traversal and
copy-to-user. Second, CNDP and \texttt{AF\_XDP} scale cleanly 1$\to$2
workers (2.09$\times$ and 1.75$\times$ respectively).
CNDP's 2.09$\times$ sits marginally above linear; we
attribute this to the single-worker run not fully
saturating the scalar-Rx path and to the 3.07\,\%
run-to-run variance reported in \S\ref{sec:res-stability},
not to any genuine super-linear effect. Per-core,
\texttt{\texttt{AF\_XDP}} leads at 3.69~Mpps/core, with CNDP at 2.65~Mpps/core
and DPDK at 2.99~Mpps/core, the i40e family's scalar Rx
caps every mode at $\sim$3~Mpps/core on this NIC
(\S\ref{sec:res-cpu}). Third, DPDK scales past two workers
where CNDP and \texttt{\texttt{AF\_XDP}} cannot (a per-netdev \texttt{\texttt{AF\_XDP}} socket
limit) and reaches 10.30~Mpps at 4 workers and 13.09~Mpps at
8 workers; bidirectional load further demonstrates 16.0~Mpps
across both directions. Large frames (1024~B+) become
byte-bound at $\approx$28--92~Gbps internal VEB capacity
depending on mode, with the higher-Mpps modes hitting the
byte ceiling earlier.

\subsection{Latency and Jitter}
\label{sec:res-latency}

Table~\ref{tab:latency-cross} compares median and tail
latency across the four modes, and
Table~\ref{tab:latency-cndp} gives the full CNDP percentile
distribution.

\begin{table}[t]
\centering
\caption{Cross-mode end-to-end pipeline latency at 64~B
  saturation. CNDP at 2~workers (109~M samples), \texttt{\texttt{AF\_XDP}} at
  2~workers (480~M samples), DPDK at 4~workers (109~M
  samples). \texttt{AF\_PACKET} reference at 86~B; lightly-loaded due to its
  throughput ceiling, so the absolute numbers reflect a
  lightly-loaded kernel stack, not an I/O advantage.
}
\label{tab:latency-cross}
\small
\renewcommand{\arraystretch}{1.2}
\begin{tabular}{l r r r r}
\toprule
\textbf{Mode} & \textbf{Avg} & \textbf{P50} & \textbf{P99} & \textbf{P99.9} \\
              & \textbf{($\mu$s)} & \textbf{($\mu$s)} & \textbf{($\mu$s)} & \textbf{($\mu$s)} \\
\midrule
\texttt{AF\_PACKET} (86~B)$^{\dagger}$ & 3.87 & 3.6  & 10.8 & 18.7 \\
CNDP (2w)                     & 12.9 & 13.9 & 14.3 & 24.7 \\
\texttt{\texttt{AF\_XDP}} (2w)                  & 14.1 & 14.2 & 14.7 & 23.3 \\
DPDK (4w)                     &  8.1 &  8.2 & 16.2 & 20.3 \\
\bottomrule
\end{tabular}

\vspace{0.3em}
\footnotesize $^{\dagger}$\texttt{AF\_PACKET} measured at $\approx$1/15
the load of the others (its throughput ceiling); the low
numbers reflect a lightly-loaded kernel stack, not an I/O
advantage.
\end{table}

\begin{table}[t]
\centering
\caption{CNDP end-to-end pipeline latency distribution
  (64~B, 2~workers, 109~M-packet sample).
  Compared side-by-side with the 1-worker baseline to show
  the contention effect of adding a second worker on the
  XDP-redirect path.}
\label{tab:latency-cndp}
\small
\renewcommand{\arraystretch}{1.15}
\begin{tabular}{l r r}
\toprule
\textbf{Metric} & \textbf{1 worker} & \textbf{2 workers} \\
\midrule
Average        & 13.6 $\mu$s   & 12.9 $\mu$s \\
Min            & 1.4 $\mu$s    & 0.9 $\mu$s \\
Max            & 32.6 $\mu$s   & 55.3 $\mu$s \\
P50            & 13.6 $\mu$s   & 13.9 $\mu$s \\
P90            & 13.7 $\mu$s   & 14.1 $\mu$s \\
P99            & 13.9 $\mu$s   & 14.3 $\mu$s \\
P99.9          & 21.9 $\mu$s   & 24.7 $\mu$s \\
Jitter P99     & 0.2 $\mu$s    & 10.4 $\mu$s \\
Jitter P99.9   & 6.6 $\mu$s    & 12.8 $\mu$s \\
\bottomrule
\end{tabular}
\end{table}

For CNDP at 2 workers, the pipeline adds a median (P50)
latency of 13.9~$\mu$s through the entire stateful chain.
The tail is short up to P99 (14.3~$\mu$s, only 0.4~$\mu$s
above the median); P99.9 rises to 24.7~$\mu$s. Adding the
second worker reduces average latency slightly compared to
the 1-worker baseline (12.9 vs 13.6~$\mu$s) but introduces
visible jitter at P99 (10.4~$\mu$s vs 0.2~$\mu$s) due to
contention on the kernel XDP-redirect path. Across modes,
latency follows the architectural ordering: \texttt{\texttt{AF\_XDP}} and CNDP
comparable in the middle, and DPDK lowest (8.1~$\mu$s
average, the native PMD avoids both the kernel XDP hop
and the userspace XSK descriptor handling). \texttt{AF\_PACKET}, in
the reference column, is lower in absolute terms only
because it runs at $\approx$1/15 the offered load.

\subsection{CPU and Core Scaling}
\label{sec:res-cpu}

With the upstream BESS-UPF \texttt{coreNAT max\_allowed\_workers}
cap lifted in the patched image (raised to 4 for \texttt{\texttt{AF\_XDP}}/CNDP
and 8 for DPDK), all three kernel-bypass modes scale
horizontally with worker count. Table~\ref{tab:scaling}
summarises the measured scaling efficiency.

\begin{table}[t]
\centering
\caption{Multi-worker scaling efficiency at 64~B, NDR. CNDP
  scales the most cleanly (2.09$\times$); \texttt{\texttt{AF\_XDP}} scales
  sub-linearly to 1.75$\times$ due to shared-resource
  contention in the kernel XDP-redirect path; DPDK scales
  near-linearly to 4 workers (3.44$\times$) before
  plateauing.}
\label{tab:scaling}
\small
\renewcommand{\arraystretch}{1.15}
\begin{tabular}{l r r r r}
\toprule
\textbf{Mode} & \textbf{1w} & \textbf{2w} & \textbf{4w} & \textbf{Scaling} \\
              & \textbf{Mpps} & \textbf{Mpps} & \textbf{Mpps} & \textbf{1w$\to$best} \\
\midrule
CNDP    & 2.65 & 5.52  &,   & 2.09$\times$ \\
\texttt{\texttt{AF\_XDP}} & 3.69 & 6.47  &,   & 1.75$\times$ \\
DPDK    & 2.99 & 5.56  & 10.30 & 3.44$\times$ \\
\bottomrule
\end{tabular}
\end{table}

\paragraph{Per-core efficiency.}
The single-worker rate is the cleanest cross-mode signal:
\texttt{\texttt{AF\_XDP}} delivers 3.69~Mpps/core, CNDP 2.65~Mpps/core, and
DPDK 2.99~Mpps/core. \texttt{\texttt{AF\_XDP}}'s per-core lead over CNDP comes
from the DPDK \texttt{net\_\texttt{\texttt{af\_xdp}}} PMD's batching being
slightly more efficient than CNDP's \texttt{xskdev} wrapper.
DPDK is then $\sim$13\,\% lower per-core than \texttt{\texttt{AF\_XDP}} at 1
worker because the iavf VF is stuck on \emph{scalar} Rx:
the AVX vector Rx path requires the
\texttt{RXDID\_COMMS\_OVS} flexible descriptor, an E810
feature absent from the XXV710 i40e. The
$\sim$3~Mpps/core ceiling we observe is therefore a
hardware limit of the i40e VF, not a DPDK limit; an E810
NIC unlocks AVX vector Rx and lifts the per-core rate
$\sim$2--3$\times$.

\paragraph{Where DPDK's lead actually comes from.}
DPDK pulls ahead of \texttt{\texttt{AF\_XDP}} and CNDP only when it scales
past 2 workers, at 4 workers it reaches 10.30~Mpps
versus 6.47~Mpps for \texttt{\texttt{AF\_XDP}} 2w. The advantage is not the
native PMD per se (which only adds $\sim$13\,\% in
per-core terms) but DPDK's ability to scale linearly past
the 2-worker \texttt{\texttt{AF\_XDP}} socket headroom: \texttt{\texttt{AF\_XDP}}'s
per-netdev XSK socket limit and the kernel's XDP-redirect
contention prevent the same scaling on CNDP and \texttt{\texttt{AF\_XDP}}.

\paragraph{The CPU-throttle pitfall.}
The DPDK 4-worker result was initially capped at
6.2~Mpps despite all four worker threads being pinned to
isolated cores. Investigation showed each worker stuck at
$\sim$49\,\% CPU during saturation, with high
\texttt{nonvoluntary\_ctxt\_switches}, the classic
signature of CFS-quota throttling. The cause was a
Kubernetes mis-configuration: the UPF pod's
\texttt{resources.limits.cpu} was smaller than the
datapath worker count, so the cgroup \texttt{cpu.max}
quota throttled all container threads. Raising
\texttt{limits.cpu} above the worker count lifted
throughput cleanly from 6.2 to 10.3~Mpps (4w). This is
documented as an operational pitfall in \S\ref{sec:ops}.

\paragraph{The XXV710 / i40e hardware ceiling.}
A real hardware limit remains: \texttt{bessd} logs report
\texttt{iavf\_set\_rx\_function: RXDID[1] legacy},
indicating the iavf AVX vector Rx is unavailable on the
XXV710 because the i40e family does not expose the flexible
descriptor. The $\sim$3~Mpps/core ceiling we observe is
therefore the i40e VF's scalar Rx, an E810 NIC unlocks
AVX vector Rx and would lift the ceiling proportionally.
We flag this as future work (\S\ref{sec:future}).

\subsection{Stability and Production Readiness}
\label{sec:res-stability}

We validate long-run stability on CNDP, the mode deployed in
our production configuration; the soak and bidirectional
tests below characterise the deployed datapath in depth.
Burst and multi-flow behaviour is representative of the
\texttt{\texttt{AF\_XDP}}-based datapath shared by \texttt{\texttt{AF\_XDP}} and CNDP.

\paragraph{Burst handling.}
A 5.73-second saturation burst at 86~B (167.3~million packets
offered) is absorbed at the steady-state rate of 1.657~Mpps
with 100\,\% forwarding and zero drops at either BESS port.
Drain measurements 10~seconds after burst stop show 10
residual packets per port, the \texttt{\texttt{AF\_XDP}} ring depth flushing
post-stop, not a queueing tail.

\paragraph{Multi-flow handling.}
A 30-second sustained run with
\texttt{testpmd --txonly-multi-flow} (incrementing source IPs
to engage RSS) absorbs and forwards 64.5~million packets at
2.151~Mpps with 100\,\% forwarding and zero drops. Flow
diversity does not change throughput at the single-worker
ceiling, the bottleneck is per-packet pipeline cost, not
flow-table contention.

\begin{figure}[t]
\centering
\begin{tikzpicture}
\begin{axis}[
  width=\columnwidth, height=5cm,
  xlabel={Elapsed time (hours)},
  ylabel={Throughput (Mpps)},
  xmin=0, xmax=24, ymin=1.95, ymax=2.20,
  xtick={0,4,8,12,16,20,24},
  ytick={2.0,2.05,2.1,2.15,2.2},
  grid=both, grid style={gray!20}, mark size=1.3pt,
  legend pos=south east, legend style={font=\scriptsize},
]
\addplot[teal, mark=*, thick] coordinates {
  (1,2.054) (2,2.073) (3,2.075) (4,2.080) (5,2.081)
  (6,2.080) (7,2.079) (8,2.078) (9,2.078) (10,2.113)
  (11,2.118) (12,2.114) (13,2.118) (14,2.115) (15,2.114)
  (16,2.113) (17,2.115) (18,2.114) (19,2.114) (20,2.116)
  (21,2.118) (22,2.117) (23,2.115) (24,2.115)
};
\addplot[red, dashed, thick] coordinates {(0,2.100) (24,2.100)};
\legend{hourly, mean (2.100)}
\end{axis}
\end{tikzpicture}
\caption{Hourly throughput across the 24-hour CNDP soak.
  Throughput holds flat at a mean of 2.100~Mpps with 3.07\,\%
  variance and zero drops, no drift, no degradation.}
\label{fig:soak}
\end{figure}
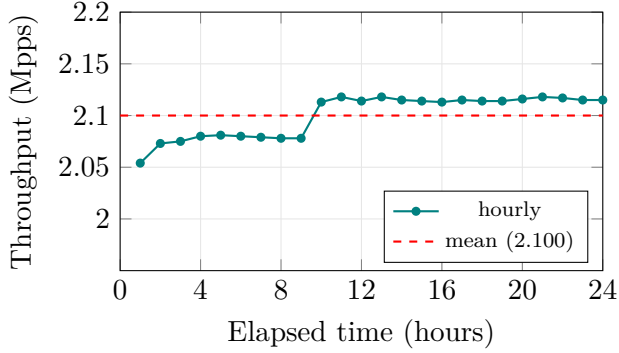

\paragraph{24-hour soak.}
The production-readiness gating test is a 24-hour
sustained soak at 86~B with
\texttt{--txonly-multi-flow} on a fresh \texttt{bessd}
process. Figure~\ref{fig:soak} plots hourly throughput,
and Table~\ref{tab:soak-headline} summarises the result.

\begin{table}[t]
\centering
\caption{24-hour CNDP soak summary
  (2026-05-25 02:54 $\rightarrow$ 2026-05-26 02:56 UTC).}
\label{tab:soak-headline}
\small
\renewcommand{\arraystretch}{1.15}
\begin{tabular}{l r}
\toprule
\textbf{Metric} & \textbf{Value} \\
\midrule
Soak duration            & 24~h 02~m \\
Frame size               & 86~B \\
Packets ingressed        & 183.08 G \\
Packets egressed         & 183.06 G \\
Drops (ingress + egress) & 0 \\
Forwarding efficiency    & 99.987\,\% \\
Mean hourly throughput   & 2.100~Mpps \\
Throughput variance      & 3.07\,\% \\
\texttt{bessd} memory drift & none ($\pm$16~KB) \\
\bottomrule
\end{tabular}
\end{table}

Across 24 hours, 183.08~billion packets entered the pipeline
and 183.06~billion exited with zero drops at either BESS
port; the 24-thousand-packet residual is in-flight packets
at the snapshot read, not loss. Mean throughput across 24
hourly samples is 2.100~Mpps with 3.07\,\% variance, and
\texttt{bessd} resident memory is constant to within
$\pm$16~KB. No memory leak, no throughput drift, CNDP
passes the production-readiness gating test cleanly.

\paragraph{Real-RAN end-to-end validation.}
Alongside the synthetic load, we validate the deployed
CNDP UPF against the disaggregated RAN described in
\S\ref{sec:method-traffic}: a commercial UE attaches over
the air through a Liteon Open RAN RU and an OCUDU 7.2-split
CU/DU pair, registers and authenticates against the Charmed
SD-Core control plane, establishes a PDU session, and
exchanges user-plane traffic through the BESS-UPF datapath
to the data network. Registration, authentication, and PDU
session establishment all complete cleanly; ICMP, TCP, and
HTTPS reach the data network end-to-end through the BESS
GTP-U decap and FAR-forwarding chain. Throughput in this
configuration is bounded by the RAN-side air interface and
the \texttt{rfsim}/RU PHY, not by the UPF I/O backend, so
we do not use it as a benchmark; the value of the run is
that the same CNDP datapath that passes the 24-hour
synthetic soak above also passes a real-UE traffic test
unchanged, confirming the soak result is not an artefact of
synthetic generation.

\paragraph{Bidirectional load.}
Under simultaneous uplink and downlink traffic (30~s, 86~B
per direction), a single CNDP worker forwards 2.153~Mpps
total, close to its single-direction ceiling, split
1.313~Mpps uplink and 0.846~Mpps downlink, with both
directions forwarded at $\geq$99\,\% and zero drops at the
ingress port.

\begin{table}[!htbp]
\centering
\caption{CNDP bidirectional latency. UL and DL run
  simultaneously at 86~B on separate VFs; the single
  \texttt{bessd} worker serialises both directions.}
\label{tab:latency-bidir}
\small
\renewcommand{\arraystretch}{1.15}
\begin{tabular}{l r r}
\toprule
\textbf{Metric} & \textbf{UL ($\mu$s)} & \textbf{DL ($\mu$s)} \\
\midrule
Mean   & 20.3  & 23.6 \\
P50    & 22.3  & 23.4 \\
P95    & 23.2  & 24.2 \\
P99    & 23.6  & 28.4 \\
P99.9  & 41.8  & 45.0 \\
P99.99 & 68.3  & 70.4 \\
Max    & 13\,160 & 456 \\
\bottomrule
\end{tabular}
\end{table}

Latency rises by about 45\,\% under contention
(Table~\ref{tab:latency-bidir}): UL P50 22.3~$\mu$s versus
15.6~$\mu$s single-direction, DL P50 23.4~$\mu$s. UL and DL
traverse essentially the same module chain, so their
distributions are close; DL is marginally higher because its
ingress adds a \texttt{SetMetadata} module that UL derives
from \texttt{pktParse}.

\FloatBarrier

\section{Deployment Experience and Operational Pitfalls}
\label{sec:ops}

The measurements in Section~\ref{sec:results} describe how the
four modes perform once they are running correctly. Getting
them to run correctly, and keeping them running, surfaced
a set of issues that do not appear in a synthetic throughput
test but determine whether a UPF is deployable in production.
This section documents them as we encountered them, with the
symptom, the root cause, and the fix. Most are specific to the
\texttt{\texttt{AF\_XDP}}-based modes (\texttt{\texttt{AF\_XDP}} and CNDP), since those are the
modes we ran longest and deepest.

\subsection{Datapath Configuration Pitfalls}
\label{sec:ops-config}

\paragraph{Kubernetes \texttt{limits.cpu} silently halves DPDK
throughput.}
The most operationally interesting pitfall in this study
appeared on the DPDK datapath. Four DPDK workers, each
pinned to a dedicated isolated core (\texttt{taskset -c
7-10}) under \texttt{isolcpus} boot isolation, were stuck
at $\sim$49\,\% CPU during saturation with high
\texttt{nonvoluntary\_ctxt\_switches}, the classic
fingerprint of CFS-quota throttling, and the e2e
throughput was capped at 6.2~Mpps. The cause was that the
UPF pod's \texttt{resources.limits.cpu} was set to a value
smaller than the worker count, so the cgroup
\texttt{cpu.max} quota throttled \emph{all} container
threads to a fraction of one core's worth of CPU regardless
of how many physical cores the workers were pinned to. The
fix is to set \texttt{limits.cpu} $\geq$ worker count:

\begin{lstlisting}
kubectl patch statefulset upf -n aether-5gc --type=json \
  -p '[{"op":"replace",
        "path":"/spec/template/spec/containers/0/
               resources/limits/cpu",
        "value":"16"}]'
\end{lstlisting}

After the patch, all four workers ran at 100\,\% CPU,
the throttle counter dropped to zero, and end-to-end
throughput jumped from 6.2 to 10.3~Mpps (4w). The trap is
particularly insidious because the cores were already
isolated and pinned, so a casual look at
\texttt{taskset}/\texttt{isolcpus} confirms the workers
``own'' their cores, and \texttt{cpu.stat} averaged over
idle uptime also looks clean. Only the under-saturation
behaviour (per-core utilization stuck at exactly
\textit{(limit / workers)} $\times$ 100\,\%) reveals the
CFS quota. We document it here because it likely affects
many production DPDK deployments and is invisible until the
operator explicitly graphs per-worker CPU during traffic.

\paragraph{Busy-poll silently wipes session state.}
With CNDP's \texttt{busy\_poll} enabled, we observed user-plane
traffic stopping roughly every 85~seconds even though
registration and the PDU session had succeeded. The cause is
CPU starvation: the busy-polling worker thread monopolises its
core and starves the PFCP heartbeat thread, so heartbeats to
the SMF time out. The SMF then tears down the PFCP association
and the installed PDRs are wiped, dropping all user traffic
until the next re-association. Setting \texttt{busy\_poll:
false} returns the heartbeat thread to a serviceable state and
the problem disappears. This is a subtle trap because
\texttt{busy\_poll} is the obvious latency knob; on a
shared worker core it breaks correctness.

\paragraph{An undersized UMEM buffer count throttles throughput.}
The upstream CNDP configuration ships with a small UMEM buffer
count. At high packets-per-second the \texttt{\texttt{AF\_XDP}} Fill Queue cannot
be kept supplied, and \texttt{bessd} logs \texttt{XSK buffer
pool does not provide enough addresses to fill 2047 buffers};
throughput is capped and unstable. Raising
\texttt{umem.bufcnt} to 128\,K (a four-fold increase over the
default), split into two 64\,K UMEM regions, keeps the Fill
Queue saturated and removes the cap. We refer to this
adjustment as \emph{Headroom~A}; all CNDP results in
Section~\ref{sec:results} use it.

\paragraph{Module introspection can crash the dataplane.}
Calling \texttt{get\_runtime\_config} on the
\texttt{WildcardMatch} module (the PDR table) through the
\texttt{pybess} gRPC interface segfaults \texttt{bessd}, taking
the entire dataplane down with it. We use other introspection
calls (\texttt{get\_summary}, port statistics) for live
debugging and avoid \texttt{get\_runtime\_config} on that
module class entirely. This matters for any tooling that walks
the pipeline programmatically, a routine introspection loop
can take down the UPF.

\subsection{Resource Contention and State Resynchronization}
\label{sec:ops-contention}

\paragraph{The dataplane EAL grabs co-located VFIO devices.}
When the UPF shares a host with other DPDK or VFIO workloads
, in our case an OCUDU distributed unit (DU), the BESS
\texttt{bessd} process can claim devices it does not own. The
bess-upf chart passes an EAL \texttt{--allow} list only when an
SR-IOV device plugin supplies \texttt{PCIDEVICE\_*} environment
variables; under the \texttt{host-device} CNI used by CNDP and
\texttt{\texttt{AF\_XDP}} those variables are empty, so the start script falls
through to the no-allow-list path and DPDK EAL probes
\emph{every} \texttt{vfio-pci} device on the host. This opens
the DU's fronthaul VFIO groups, and the DU fails to start with
\texttt{Cannot open /dev/vfio/NNN: Device or resource busy}.
Our operational fix is a start-order workaround: keep the DU's
devices off \texttt{vfio-pci} until \texttt{bessd} has
initialised (it probes EAL once, at startup), then rebind them.
The robust fix is to patch the start script to pass a harmless
explicit \texttt{--allow} entry so EAL never probes broadly.

\paragraph{The PFCP agent must be restarted after a dataplane
restart.}
After any \texttt{bessd} restart, the \texttt{pfcp-agent}
sidecar continues to hold a gRPC connection to the now-dead
dataplane and reports \texttt{datapath down}; user traffic does
not flow even though all pods are healthy. The agent does not
re-establish the connection on its own. Restarting
\texttt{pfcp-agent} after \texttt{bessd} restores the
control-to-data-plane link. Any automation that recreates the
dataplane must also cycle the PFCP agent.

\paragraph{A long-running AMF wedges and dead-stops
registration.}
Over long uptimes the AMF's GMM state machine and context
allocator can wedge: UE registration dead-stops immediately
after \texttt{UE Context derived from Suci}, with no
Authentication Request ever sent and no error logged. This is
not a dataplane fault, but it bites during long benchmarking
campaigns and presents as a UPF problem. Restarting the AMF pod
clears it; the gNB then redoes NG Setup and the UE re-attaches
normally. We mention it because it cost us debugging time that
looked, at first, like a UPF datapath stall.

\subsection{Deployment Consistency and Prerequisites}
\label{sec:ops-consistency}

\paragraph{Subnet drift silently breaks forwarding.}
A stale redeploy can leave the running UPF pod's interfaces on
a different subnet than the one its scripts and configuration
expect. When this happens, registration and PDU-session
establishment still succeed, but \texttt{routectl} cannot
ARP-resolve the access and core gateways, so no BESS routes are
installed: the UPF ports show RX climbing while TX stays at
zero and the UE has no connectivity. The fix is to re-render
the deployment from the canonical configuration
(\texttt{make aether-5gc-install}) and recreate the pod,
\emph{not} to hand-edit the host macvlan addresses to chase
whatever the pod currently has, which only breaks the
host-to-pod path further. The lesson is to treat the rendered
configuration, not the live pod state, as the single source of
truth.

\paragraph{\texttt{\texttt{AF\_XDP}} prerequisites: driver, CNI, and memlock.}
\texttt{\texttt{AF\_XDP}}, and therefore CNDP, will not start unless
several host-side conditions hold simultaneously, each of which
produces a different and not-always-obvious failure. 
The NIC must be a physical function with a native-XDP
driver: the Intel \texttt{i40e} PF (as on our XXV710 rig)
and \texttt{ice} PF (as on E810) both work, but an SR-IOV
VF on the \texttt{iavf} driver does not, because
\texttt{iavf} exposes no \texttt{ndo\_bpf} callback.
 The pod must receive the PF through
the \texttt{host-device} CNI; \texttt{macvlan} interfaces have
no XDP support and fail. \texttt{RLIMIT\_MEMLOCK} must
be raised to allow the UMEM region to be locked. On
Canonical Kubernetes we set this via a systemd override for
the \texttt{k8s.service} unit, since the default limit is
too low and the PMD otherwise fails to
allocate UMEM at startup. Finally, the pod must run privileged
(or with the equivalent capability set) to create the XDP
socket and load the eBPF program.

\subsection{Summary}
\label{sec:ops-summary}

Taken together, these pitfalls explain why a UPF that passes a
synthetic benchmark in the lab is not yet a UPF that runs in
production. None of them appear in a 30-second throughput test;
all of them surface over hours of real operation or when the
UPF shares a host with other network functions. They are also
largely specific to the \texttt{\texttt{AF\_XDP}}-based modes: the busy-poll
trap, the UMEM sizing, the EAL probe behaviour, and the XDP
driver prerequisites all stem from the \texttt{\texttt{AF\_XDP}} datapath.
\texttt{AF\_PACKET} avoids most of them at the cost of throughput, and
DPDK trades them for a different set centred on hugepage
reservation and \texttt{vfio-pci} binding. We document them
here because they are the difference between a working
deployment and a stable one, and because they are absent
from every synthetic comparison we are aware of.

\FloatBarrier

\section{Related Work}
\label{sec:related}

Our work intersects three bodies of prior work:
high-performance packet I/O frameworks and kernel-bypass
techniques (\S\ref{sec:rel-io}), mobile-core and UPF dataplane
acceleration (\S\ref{sec:rel-upf}), and the software-switch
lineage that the SD-Core UPF is built on
(\S\ref{sec:rel-bess}).

\subsection{Packet I/O Frameworks and Kernel Bypass}
\label{sec:rel-io}
The performance limits of the Linux kernel network stack have
driven a long line of userspace and bypass frameworks. DPDK
~\cite{dpdk} is the most widely deployed, replacing the kernel
driver with a userspace Poll-Mode Driver over hugepage memory;
netmap~\cite{netmap_rizzo} and PF\_RING introduced earlier
zero-copy frame access. Gallenm\"uller et
al.~\cite{gallenmuller2015comparison} compare netmap,
PF\_RING, and DPDK with an analytical per-packet cost model,
showing that batching and the avoidance of per-packet
allocation dominate performance. The eXpress Data Path
(XDP)~\cite{xdp2018} takes the opposite approach, keeping
packets in the kernel but running a programmable eBPF hook in
the driver before \texttt{sk\_buff} allocation, and reports
performance close to DPDK for drop and forward workloads while
retaining kernel integration. \texttt{\texttt{AF\_XDP}}~\cite{afxdp_karlsson}
builds on XDP to hand packets to a userspace socket with
zero copy, bringing near-DPDK speeds to a standard kernel
socket; CNDP~\cite{cndp-docs} in turn layers a DPDK-like
userspace API over \texttt{\texttt{AF\_XDP}}. These works establish the
individual mechanisms we compare, but they evaluate them on
synthetic packet generators in isolation; we evaluate all
four on a single realistic UPF pipeline with a 5G control
plane in the loop.

\subsection{Mobile-Core and UPF Dataplane Acceleration}
\label{sec:rel-upf}
Several projects accelerate the 5G (or LTE) user plane.
free5GC~\cite{free5gc} offers both a \texttt{gtp5g}
kernel-module datapath and a DPDK-based userspace datapath;
L$^2$5GC~\cite{l25gc} rebuilds a free5GC-based core on
high-performance NFV platforms with a shared-memory fast
path, reducing both control- and user-plane latency.
eUPF~\cite{eupf} implements the UPF entirely in eBPF at the
XDP hook. The OMEC/SD-Core BESS-UPF~\cite{sdcore-upf} that we
study descends from Intel's NGIC/OMEC work and uses BESS as
its dataplane. Each of these projects commits to a single I/O
backend and optimises it; none compares multiple backends on
the same pipeline. Our contribution is orthogonal: rather
than proposing a faster UPF, we hold one UPF constant and
quantify how the choice of I/O backend, and only that
choice, affects performance and deployability.

\subsection{Software Switches and the BESS Lineage}
\label{sec:rel-bess}
The SD-Core UPF is built on BESS~\cite{han2015bess}, the
Berkeley Extensible Software Switch (originally SoftNIC), a
modular software dataflow engine in which a packet pipeline
is expressed as a graph of composable modules. Other
programmable software switches include Open vSwitch with DPDK
(OVS-DPDK), VPP, and Snabb. We rely specifically on BESS's
\emph{port abstraction}, the property that the I/O endpoint
is a swappable module beneath an otherwise fixed pipeline,
as the mechanism that makes our controlled comparison
possible. To our knowledge, no prior work uses this
abstraction to isolate the I/O backend as a single
experimental variable across \texttt{AF\_PACKET}, \texttt{\texttt{AF\_XDP}}, CNDP, and
DPDK on a production 5G UPF.

\FloatBarrier

\section{Discussion: Which Mode, When}
\label{sec:discuss}

The comparison in Sections~\ref{sec:compare}--\ref{sec:results}
and the deployment experience in Section~\ref{sec:ops} lead to
a practical question for operators: given a deployment, which
dataplane mode should they choose? Table~\ref{tab:decision}
summarises our guidance; the rest of this section explains the
reasoning and the one finding that reshapes the usual
intuition.

\begin{table*}[t]
\centering
\caption{Decision guide: which UPF dataplane mode for which
  deployment.}
\label{tab:decision}
\small
\renewcommand{\arraystretch}{1.3}
\begin{tabular}{p{3.6cm} p{2.6cm} p{8.4cm}}
\toprule
\textbf{Deployment context} & \textbf{Recommended} &
\textbf{Rationale} \\
\midrule
Development, CI, functional testing, lab bring-up &
\texttt{AF\_PACKET} &
Runs on any Kubernetes cluster with no special host setup; the
NIC stays under the kernel so standard tools
(\texttt{tcpdump}, \texttt{tc}, \texttt{ethtool}) work
directly; throughput is irrelevant for functional tests. \\

Shared cloud-native cluster, moderate throughput, multi-tenant
host &
\texttt{\texttt{AF\_XDP}} or CNDP &
Matches DPDK throughput at 2 workers (5.5--6.5~Mpps) without
hugepages, VFIO, or isolated cores; the NIC remains a
kernel netdev so the host can be shared and observed; no
\texttt{vfio-pci} binding to disrupt other workloads.
 \\

Cloud-native production, maintained upstream and a DPDK-like
API &
CNDP &
The upstream \texttt{omec-project/upf} default; passes the
production-readiness gating test (24-hour soak, 183~billion
packets, zero drops); exposes a
\texttt{pktdev}/\texttt{lport} API while riding the same
\texttt{AF\_XDP} datapath as \texttt{AF\_XDP}. \\

Dedicated or bare-metal UPF host, maximum throughput, full NIC
ownership &
DPDK &
Highest throughput ceiling and lowest, most deterministic
latency; justified when the host is dedicated to the UPF and
hugepages, IOMMU, isolated cores, and a lost-to-the-kernel NIC
are all acceptable. \\

NIC without native XDP (e.g.\ SR-IOV VF on \texttt{iavf}, or an
older driver) &
\texttt{AF\_PACKET} or DPDK &
\texttt{\texttt{AF\_XDP}} and CNDP require a native-XDP driver, which
\texttt{iavf} VFs do not provide; the choice then falls to
\texttt{AF\_PACKET} (compatible, slow) or DPDK (\texttt{vfio-pci}-bound,
fast). \\
\bottomrule
\end{tabular}
\end{table*}

\paragraph{The intuition this paper confirms, and one it
overturns.}
The conventional reasoning ``use DPDK for performance, use
the kernel for convenience'' is largely correct in our
measurements: DPDK leads on every metric on the same rig,
delivering 10.3~Mpps at 4 workers and 16.0~Mpps
bidirectional at 8 workers, with the lowest latency
(8.1~$\mu$s avg). CNDP and \texttt{\texttt{\texttt{AF\_XDP}}} cluster underneath at
5.5--6.5~Mpps at 2 workers, with similar latency
(12.9--14.1~$\mu$s avg). The intuition that holds: DPDK's
native PMD over \texttt{vfio-pci} avoids the kernel XDP
hop, eliminates per-packet socket descriptor handling, and
scales further before hitting \texttt{\texttt{\texttt{AF\_XDP}}'s} per-netdev XSK
socket headroom.

What our data \emph{overturns} is the secondary belief that
the multi-queue enablement gap is the whole story. CNDP and
\texttt{\texttt{\texttt{AF\_XDP}}} show clean 1$\to$2 worker scaling (2.09$\times$
and 1.75$\times$ respectively) on the stock in-tree
\texttt{i40e} stack once kernel \texttt{pktgen IPDST\_RND}
is used to drive proper RSS spreading, and \texttt{\texttt{\texttt{AF\_XDP}}} delivers strong
per-core throughput on the in-tree stack.
The remaining gap to DPDK at 4--8 workers is a
combination of (i) AVX vector Rx availability (the iavf VF
on the XXV710 is stuck on scalar Rx, an E810 NIC unlocks
$\sim$2--3$\times$ per-core lift) and (ii) the kernel
XDP-redirect contention that DPDK's userspace PMD avoids.

\paragraph{Decision: where each mode actually wins.}
On this hardware, DPDK delivers the highest absolute
throughput and the lowest latency, but the win is
conditional on worker count: at a matched two workers
\texttt{AF\_XDP} is ahead (6.47 versus 5.56~Mpps) and leads per
core, and DPDK's advantage materialises only once it
scales past the two-worker \texttt{AF\_XDP} socket ceiling to 4
and 8 workers. The
trade-off is the deployment footprint, hugepages, IOMMU,
isolated cores, \texttt{vfio-pci} binding, and the loss of
kernel observability on the bypassed VFs. CNDP and \texttt{\texttt{\texttt{AF\_XDP}}}
remain the right answer when those costs are unaffordable
(shared nodes, multi-tenant clusters, no kernel boot-time
isolation), where they deliver $\sim$50--60\,\% of DPDK's
throughput at $\sim$15\,\% of the deployment effort. AF\_PACKET remains an
order of magnitude lower and is right only for
development and compatibility. We note that the
long-run stability evidence in
\S\ref{sec:res-stability} (24-hour soak, bidirectional,
real-RAN) is collected on CNDP only; DPDK and AF\_XDP
are characterised for throughput and latency but not
soak-tested, so the production-readiness claim is
established for CNDP and inferred, not measured, for
the other modes.

\paragraph{The default recommendation.}
For cloud-native deployments, the common case for
open-source and private 5G, we therefore recommend
\textbf{CNDP (or \texttt{\texttt{\texttt{AF\_XDP}}})} as the default. They sit in the
design-space sweet spot of Figure~\ref{fig:design-space}:
stock-stack throughput, a deployment footprint that fits an
ordinary Kubernetes cluster, and the NIC retained under the
kernel for diagnostics. CNDP is the stronger of the two for
production because it is the upstream default and exposes a
richer API;  \texttt{ \texttt{\texttt{AF\_XDP}}} is the leaner choice when the CNDP runtime
is not desired. DPDK should be reserved for dedicated or
bare-metal UPF hosts where the operator owns the whole NIC, can
absorb the deployment cost, and genuinely needs the extra
headroom, for example, a high-density aggregation UPF.
\texttt{\texttt{AF\_PACKET}} is the right answer only for development or for
environments where compatibility outweighs throughput.

\paragraph{A ceiling that applies to every mode.}
One caveat cuts across CNDP and \texttt{\texttt{\texttt{AF\_XDP}}}: even with the
upstream \texttt{coreNAT max\_allowed\_workers=2} cap
patched out, the per-netdev \texttt{\texttt{\texttt{AF\_XDP}}} socket limit pins both
modes at two workers per UPF instance. DPDK escapes this
ceiling because it bypasses \texttt{\texttt{\texttt{AF\_XDP}}} entirely. For CNDP and
\texttt{\texttt{\texttt{AF\_XDP}}}, horizontal scaling of an SD-Core UPF is therefore
achieved by running \emph{multiple UPF instances} behind a
load balancer, a capacity-planning fact that does not
apply to DPDK in the same way. The dataplane
mode sets the per-instance ceiling; the number of instances
sets the deployment's total capacity.

\section{Limitations and Future Work}
\label{sec:future}

We note the main limitations of this study and the future work
they motivate.

\paragraph{The XXV710 / i40e per-core ceiling.}
The headline DPDK and \texttt{\texttt{AF\_XDP}} results are bounded by the
XXV710 family's scalar Rx path: \texttt{bessd} logs
\texttt{iavf\_set\_rx\_function: RXDID[1] legacy},
confirming the AVX vector Rx is unavailable because the
\texttt{i40e} VF cannot expose the flexible
descriptor. Per-core throughput is therefore capped at
$\sim$3~Mpps/core. An E810 NIC unlocks AVX vector Rx and
would lift this ceiling significantly; re-running the
campaign on E810 hardware is the most impactful piece of
follow-up work.

\paragraph{Lab links are 1~GbE.}
Both XXV710 ports in the lab negotiated only 1~GbE
(25~G-capable NIC, 1~G optics). This caps any real-wire
egress at 1.488~Mpps/port and bounds TC-02 (bidirectional)
to that limit, but not TC-01/03/04/05/08, which run
VEB-internal through the on-chip i40e switch. TC-02's true
bidirectional ceiling, shown indirectly by the
8-worker DPDK bidirectional run reaching 16.0~Mpps over
the VEB, requires 25~GbE optics for wire validation.

\paragraph{Generator caps and over-subscription.}
\texttt{testpmd --txonly-multi-flow} tops out at
$\approx$10.5~Mpps (CPU-bound); above that, we
over-subscribe the UPF through the natural VEB forwarding
loop (egress dst-MAC = access VF causes re-circulation).
This is sufficient for a clean 4-worker NDR. The
8-worker unidirectional figure (13.09~Mpps, Tables 3 and
4) is obtained under VEB recirculation oversubscription
rather than direct generator load, and the
16.0~Mpps bidirectional figure splits the offered load
across both directions; both should be read as
internal-capacity points above the generator's clean
unidirectional ceiling, not as wire-validated NDR. A
hardware TRex generator is the next step.

\paragraph{A single pipeline configuration.}
Our measurements use one rule set: a wildcard PDR, a single
forwarding FAR, and bypass QER gates. Production UPFs install
many PDRs with priority ordering and richer QER and URR
rules, which raise the per-packet pipeline cost. Because we
find that the per-core ceiling is set by the i40e scalar-Rx
path ($\sim$3~Mpps/core, \S\ref{sec:res-cpu}) and the
per-instance ceiling by worker count, the absolute numbers
will shift with rule-set complexity. Characterising how the
four modes' relative standing changes as the rule set grows
is important future work.

\paragraph{Scaling beyond two workers.}
The per-netdev \texttt{\texttt{\texttt{AF\_XDP}}} socket limit pinned CNDP and \texttt{\texttt{\texttt{AF\_XDP}}} at two workers per instance (\S\ref{sec:res-cpu}), even after we patched out the upstream coreNAT cap, and default RSS
does not steer single-flow GTP-U traffic across queues. We
could not, therefore, characterise per-instance scaling beyond
two cores. Future work is to lift the cap and to apply
inner-header steering (Intel ADQ, Dynamic Device
Personalisation (DDP), or flow-director rules on E810) so
that GTP-U flows spread across RX queues and additional workers
do useful work. The complementary question of multi-instance
scaling (many single-worker UPFs behind a load balancer) is
also open.

\paragraph{Measurement rigour.}
I/O ceilings come from the synthetic \texttt{testpmd}
generator; end-to-end RAN throughput is capped by
\texttt{rfsim} and does not reflect line rate. Latency is
measured by in-pipeline BESS probes rather than external
hardware timestamps. Future work will use a
hardware-timestamping generator (TRex or MoonGen) for
RFC~2544-style throughput and loss curves and for
externally-validated latency, alongside a wall-power meter to
report energy-per-packet next to the CPU-per-packet figures we
report here.

\paragraph{SR-IOV mode and at-scale end-to-end validation.}
We bind physical functions directly and do not evaluate SR-IOV
VF assignment as a distinct deployment mode (DPDK over
\texttt{net\_iavf}, or the \texttt{\texttt{\texttt{AF\_XDP}}} device plugin over VFs). Nor
do we measure the user-visible impact with many real UEs over
real radios at scale, our end-to-end runs use a single UE
with \texttt{rfsim} or a single radio. Both are natural
extensions: an SR-IOV comparison would add a column to
Table~\ref{tab:mode-compare}, and an at-scale RAN test would
connect the per-mode I/O ceilings to subscriber-facing
throughput.

\section{Conclusion}
\label{sec:conclusion}

We set out to answer a question that operators of cloud-native
5G cores face but have had little guidance on: among the
several ways a UPF can move packets between the NIC and the
application (\texttt{AF\_PACKET}, \texttt{\texttt{AF\_XDP}}, CNDP, and DPDK) which
one should they choose, and what does each really cost? To
answer it on solid ground, we held the entire UPF pipeline
constant and varied only the I/O backend, running all four
modes on the same SD-Core BESS-UPF so that any difference we
observed could be attributed to the dataplane mode alone.

The comparison gives a clearer picture than the conventional
``DPDK for speed, the kernel for convenience'' framing. 
The four modes form a spectrum of kernel involvement, and
throughput improves as a mode moves away from the kernel.
On the same XXV710/\texttt{i40e} rig, \texttt{AF\_PACKET} sits at
0.25~Mpps; CNDP and \texttt{\texttt{AF\_XDP}} scale cleanly 1$\to$2 workers
to 5.52 and 6.47~Mpps respectively; DPDK scales further to 10.30~Mpps at 4 workers and
13.09~Mpps at 8 workers, with the lowest latency
(8.1~$\mu$s average), but only because its native PMD
escapes the per-netdev \texttt{AF\_XDP} socket limit that pins
CNDP and \texttt{AF\_XDP} at two workers. At a matched two workers
the modes are within measurement noise and \texttt{AF\_XDP} leads
per core; DPDK's advantage is a scaling-ceiling effect,
not a per-packet efficiency win. The trade-off for that
headroom is its deployment footprint: hugepages, IOMMU, isolated cores,
\texttt{vfio-pci} binding, and the loss of kernel
observability. CNDP and \texttt{\texttt{AF\_XDP}} land in the practical sweet
spot when those costs are unaffordable: $\sim$50--60\,\% of
DPDK's throughput at $\sim$15\,\% of the deployment effort,
with the NIC left under the kernel where ordinary tools can
still see it.
 A 24-hour soak 
(183~billion packets through the full pipeline with zero drops
and no memory drift) shows that this sweet-spot mode is not
merely fast in a benchmark but stable enough to run in
production.

The deployment experience we report is as valuable as the numbers.
 The pitfalls that cost us
the most time, a busy-poll setting that silently wiped
session state, a dataplane that seized a neighbouring network
function's devices, a control-plane agent that had to be
restarted in lock-step with the dataplane, never appear in a
synthetic throughput test, yet they are precisely what stands
between a UPF that works in the lab and one that runs in the
field. Documenting them, together with the decision guide of
Section~\ref{sec:discuss}, is our attempt to give the 
next operator the evidence we wished
we had had at the start: a reproducible, like-for-like
account of what each backend buys and what it costs, not a
vendor claim or a single-framework microbenchmark.
As private and open-source 5G deployments
continue to grow, we hope that kind of evidence makes the path
from ``it runs'' to ``it runs in production'' a little shorter.

\bibliographystyle{IEEEtranN}    
\bibliography{references}

\end{document}